\documentclass[aps,pra,reprint,groupedaddress,longbibliography]{revtex4-1}
\pdfoutput=1
\usepackage{graphicx,color}
\usepackage{amsmath, amssymb}
\usepackage{longtable}
\usepackage{natbib}

\begin{document}

\title{
Giant exchange interaction in mixed lanthanides 
}

\author{Veacheslav Vieru}
\author{Naoya Iwahara}
\author{Liviu Ungur}
\author{Liviu F. Chibotaru}
\affiliation{Theory of Nanomaterials Group, 
Katholieke Universiteit Leuven, 
Celestijnenlaan 200F, B-3001 Leuven, Belgium}
\date{\today}

\begin{abstract}
Combining strong magnetic anisotropy with strong exchange interaction is a long standing goal in the design of quantum magnets. 
The lanthanide complexes, while exhibiting a very strong ionic anisotropy, usually display a weak exchange coupling, 
amounting to only a few wavenumbers. 
Recently, an isostructural series of mixed 
Ln$^{3+}$-N$_2^{3-}$-Ln$^{3+}$ (Ln $=$ Gd, Tb, Dy, Ho, Er)
have been reported, in which the exchange splitting is estimated to reach hundreds wavenumbers. 
The microscopic mechanism governing the unusual exchange interaction in these compounds 
is revealed here by combining detailed modeling with 
density-functional theory
and {\it ab initio} calculations. 
We find it to be basically kinetic and highly complex, involving non-negligible contributions 
up to seventh power of total angular momentum of each 
lanthanide site. 
The performed analysis also elucidates the origin of magnetization blocking in these compounds. 
Contrary to general expectations the latter is not always favored by strong exchange interaction.
\end{abstract}

\maketitle
%
%

%
%
\section*{Introduction}
The effects of strong magnetic anisotropy, traditionally investigated in magnetic insulators, especially, 
in $f$-electron systems \cite{Santini2009, Zvezdin1985, Gingras2014}, recently attracted renewed interest in connection 
with molecular magnetic materials \cite{Gatteschi2006}. 
The investigation of molecular nanomagnets gave birth to new objects 
such as single-molecule magnets (SMMs) \cite{Sessoli1993, Christou2000} and single-chain magnets \cite{Coulon2006}, 
and initiated studies in the domain of molecular spintronics \cite{Heersche2006, Bogani2008} 
and quantum computation \cite{Leuenberger2001, Timco2009, Aromi2012}. 
Among them, in the last years the accent moved towards lanthanide complexes which have already demonstrated 
several exciting properties \cite{Ishikawa2004, Rinehart2011, Woodruff2013, Ungur2014, LayfieldMurugesu2015, Demir2015}.  

The key feature of lanthanide ions in materials is their strong magnetic anisotropy caused 
by strong spin-orbit coupling effects \cite{Dieke1967}, which often leads to highly 
axial ground and low-lying excited doublet states even in the lack of axial symmetry \cite{Ungur2011}. 
Due to small radius of electronic $f$-shells, the exchange interaction in lanthanide complexes is much weaker than 
the crystal-field splitting on lanthanide ions \cite{Chibotaru2015}. 
As a result, only individual doublet states on lanthanide sites, described by pseudospins $\tilde{s}=1/2$, 
participate in the magnetic interaction. 
The latter is described by a Hamiltonian bilinear in pseudospins ($\tilde{s}_1$ and $\tilde{s}_2$) 
in the case of two interacting lanthanide ions, or a pseudospin ($\tilde{s}_1$) and a true spin ($S_2$) 
in the case of a lanthanide ion interacting with a transition metal or a radical
when the spin-orbit coupling in the second site is negligible.
For strongly axial doublet states on the lanthanide sites 
(Ln)
these Hamiltonians basically become of Ising type \cite{Chibotaru2008}: 
\begin{eqnarray}
 \hat{H}_{\rm Ln-Ln} &=& - \mathcal{J} \tilde{s}_{1z_1} \tilde{s}_{2z_2}, \nonumber\\
 \hat{H}_{{\rm Ln-}S} &=& - \mathcal{J} \tilde{s}_{1z_1} \hat{S}_{2z_2}, 
\label{Eq:H_Ising}
\end{eqnarray}
either collinear ($z_1 \parallel z_2$) or non-collinear ($z_1 \nparallel z_2$) depending on 
geometry \cite{Chibotaru2015} and details of interaction. \cite{Chibotaru2015Ising}
The exchange parameter is contributed by magnetic dipolar and exchange interaction between the sites, $ \mathcal{J} = \mathcal{J}_{\text{dip}} + \mathcal{J}_{\text{exch}}$, the former being usually stronger in net lanthanide complexes \cite{Chibotaru2015}.  

\begin{figure*}[tb]
\begin{center}
 \includegraphics[bb=0 0 5543 2246, width=16cm]{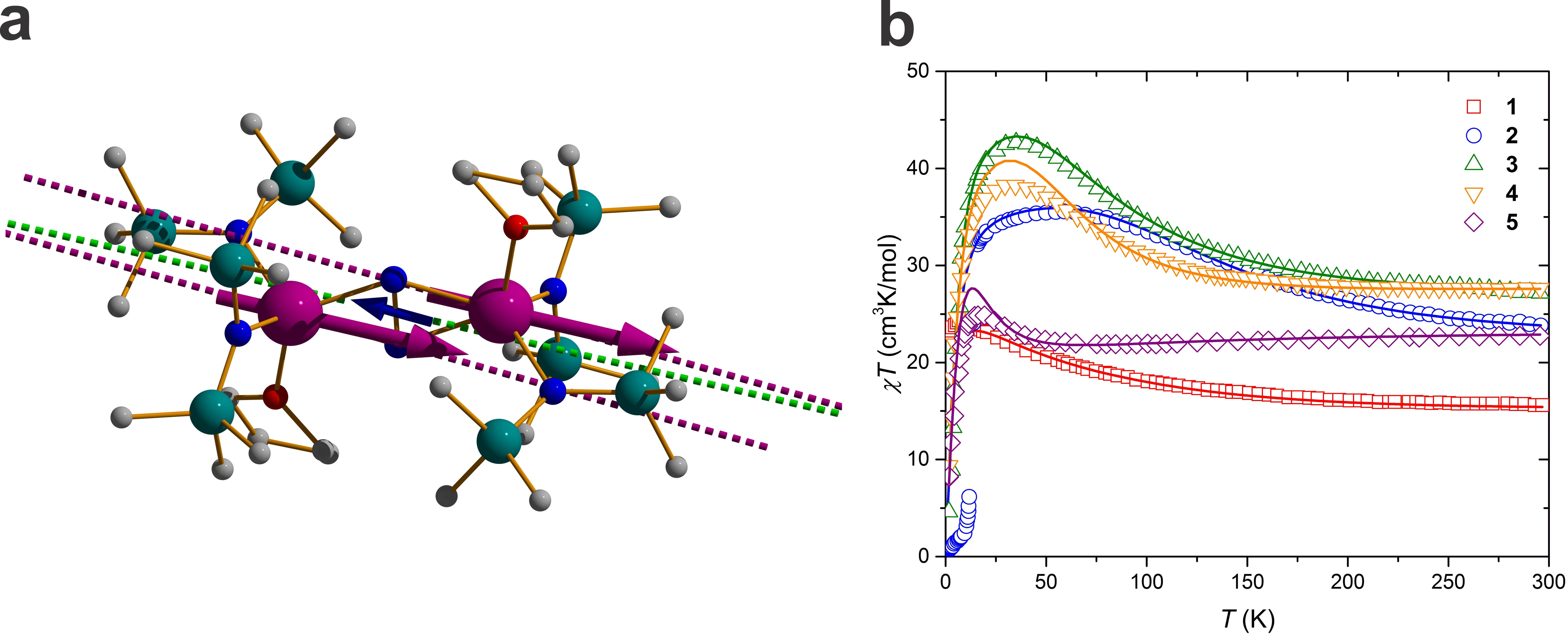}
\end{center}
\caption{
Molecular structure of Tb complex {\bf 2} and magnetic susceptibility in the series {\bf 1}-{\bf 5}. 
{\bf a}, Colors' legend for the balls: violet, Tb; blue, N; red, O; green, Si; grey, C. 
The hydrogen atoms are omitted for clarity. 
The violet dashed lines show the orientation of the main anisotropy axes of Tb ions 
in their ground doublet state, whereas the green dashed line shows the orientation 
of the main anisotropy axis in the ground exchange Kramers doublet. 
The violet arrows show the orientation of the local magnetic moments on Tb ions, and the blue arrow on the radical,
in the ground exchange Kramers doublet. 
{\bf b}, Experimental (symbols) and {\em ab initio} calculated (lines) temperature-dependent powder magnetic susceptibility 
($\chi$) for {\bf 1}-{\bf 5}. 
The experimental data were upscaled by 3, 3, 1 \% for {\bf 2}, {\bf 3}, and {\bf 5},
respectively, and were downscaled by 2 \% for {\bf 4}.
The magnetic susceptibility curves were calculated following the way they have been measured \cite{Rinehart2011,Rinehart2011JACS},
as $M(\mathbf{H},T)/H$ at $H$= 1 T, 
averaged over all directions of magnetic field $\mathbf{H}$ relative to molecular frame. 
For the computational methodology of the magnetic axes and $\chi T$, see Refs. \cite{Chibotaru2012} and \cite{Ungur2015},
respectively.
}
\label{Fig1}
\end{figure*}

This paradigm was recently challenged 
by a series of N$_2^{3-}$-radical bridged dilanthanide complexes 
[K(18-crown-6)]\{[(Me$_3$Si)$_2$N](THF)Ln\}$_2$($\mu$-$\eta^2$:$\eta^2$-N$_2$) 
(Ln = Gd ({\bf 1}), Tb ({\bf 2}), Dy ({\bf 3}), Ho ({\bf 4}), Er ({\bf 5}), 
THF = tetrahydrofuran), shown in Fig. \ref{Fig1}a \cite{Rinehart2011, Rinehart2011JACS}. 
In some of these compounds the exchange interaction was found to be two orders of magnitude stronger than 
in any known lanthanide system. 
This is of the same order of magnitude as the crystal-field splitting of $J$-multiplets on the lanthanide sites, implying that the picture of exchange interaction involving individual crystal-field doublets, Eq. (\ref{Eq:H_Ising}), is no longer valid for these compounds.
Moreover, the terbium complex from this series exhibits a magnetic hysteresis at 14 K and a 100 s blocking time at 13.9 K (one of the highest blocking temperatures among existing SMMs \cite{Rinehart2011JACS}), suggesting a possible implication of the giant exchange interaction in this SMM behavior. 

The purpose of the present work is to reveal the mechanism of giant exchange interaction
and the origin of the magnetization blocking of the series of the complexes
based on adequate theoretical treatment. 
We apply an approach combining {\it ab initio} and 
density-functional theory (DFT)
calculations with microscopic model description to unravel the nature of this exchange interaction. We also elucidate the origin of blocking barriers in these compounds and discuss the effect of strength of exchange interaction on magnetization blocking in strongly anisotropic complexes. 

\section*{Results}
\subsection*{Origin of giant exchange interaction}
To understand the origin of such strong exchange interaction, we consider 
the simplest complex of the series, the gadolinium one. 
In this system
the isotropic spins of Gd$^{3+}$ ions ($S_{\text{Gd}}=7/2$) interact with the radical spin of the N$_2^{3-}$ bridge ($S_{\text{N}_2}=1/2$) via Heisenberg exchange interaction,
$\hat{H}_\text{ex} = \sum_{i=1,2} -2\mathcal{J}_\text{Heis} \hat{\mathbf{S}}_{\text{Gd}(i)} \cdot\hat{\mathbf{S}}_{\text{N}_2}$,
described by a single parameter $\mathcal{J}_\text{Heis}$ due to the inversion symmetry of the complex (Fig. \ref{Fig1}a).
Broken-symmetry 
DFT 
calculations 
\cite{Soda2000}
give the value $\mathcal{J}_\text{Heis} = -21.4$ cm$^{-1}$ in close agreement with the experimental one, 
$\mathcal{J}_\text{Heis}=-27$ cm$^{-1}$ \cite{Rinehart2011}, and the previous DFT calculations \cite{Rajeshkumar2012, Zhang2013}. 

To get insight into the mechanism responsible for the obtained huge value of $\mathcal{J}_\text{Heis}$, 
we projected a series of DFT calculations into the effective tight-binding and Hubbard models 
acting in the space of interacting magnetic orbitals of two Gd ions and the radical
(see the Supplemental Material for details).
Because of the $D_{2h}$ symmetry of the exchange core (Fig. \ref{Fig2}a), the antibonding $\pi^*$ orbital accommodating the unpaired electron of N$_2^{3-}$ radical overlaps with only one of the $4f$ orbitals on each Ln site, the $xyz$ one (Fig. \ref{Fig2}b).
The corresponding transfer parameter $t$ was derived for the Gd complex as $t =$ 1407 cm$^{-1}$. 
The value of $t$ is obtained large because the radical's magnetic orbital $\pi^*$ resides 
on nearest-neighbor atoms (nitrogens) to both lanthanides. 
Most important, this orbital is found to lie 
higher than the $4f_{xyz}$ orbitals by as much as $\Delta =  5.2 \times 10^4$ cm$^{-1}$ (Fig. \ref{Fig2}b). 
Because of this huge energy gap, small electron promotion energy is expected 
for the electron transfer from the $\pi^*$ to the $4f_{xyz}$ orbitals: 
the Coulomb repulsion energy between the transferred electron and the $f$ electrons is cancelled at large extent by $\Delta$.
On the other hand, because of the same large gap $\Delta$, 
the promotion energy of electron transfer from $4f$ to $\pi^*$ orbital 
is at least one order of magnitude larger. 
Therefore, the contribution of this process to the exchange coupling 
can be neglected.
Indeed, 
our analysis using the Hubbard model gives the experimental $\mathcal{J}_\text{Heis}$
for the Gd complex
with (averaged) promotion energy of 
$\bar{U}= 8872$ cm$^{-1}$,  
a value many times smaller than typical ``Hubbard $U$'' in metal complexes \cite{vanderMarel1988}.
Taking into account only the dominant virtual electron transfer, 
$(4f)^7 (\pi^*)^1 \rightarrow (4f)^{8} (\pi^*)^0 \rightarrow (4f)^7 (\pi^*)^1$, 
the kinetic contribution 
to the Gd$^{3+}$-N$_2^{3-}$ exchange parameter
is written in a good approximation as $-2t^2 / \bar{U}$ \cite{Anderson1959, Anderson1963}.

Compared to this mechanism, the other contributions  
such as the direct exchange, 
the delocalization of unpaired electron of N$_2^{3-}$ 
into the empty $5d$ orbitals of Gd$^{3+}$
(Goodenough's mechanism \cite{Goodenough1963}), 
the spin polarization and 
the magnetic dipolar interaction between Gd$^{3+}$ ions are expected to be 1 - 2 orders of magnitude smaller. 
The reason is that all these contributions are expected to be of the same order of magnitude as in other 
lanthanide-radical compounds.
Indeed, the direct exchange integral depends only on the shape of the $4f$ an radical's orbitals, which is not expected to be much different from other complexes. 
The Goodenough's contribution arises from higher (third) order of the perturbation theory compared to the usual kinetic exchange, and involves the excitation energy into a higher
$5d$ orbital on the Ln site.
Both these contributions are usually neglected unless the conventional kinetic exchange appears to be small \cite{Anderson1959, Anderson1963}.
The spin polarization mechanism starts to play a role when the ligand bridging the magnetic centers contains a spectrum of low-lying orbital excitations, which is certainly not the case of N$_2^{3-}$. 
As for magnetic dipolar interaction, it is estimated for Gd$^{3+}$-N$_2^{3-}$ to be $\sim$ 0.25 cm$^{-1}$.  

The same physical situation is realized in the other complexes of the series. 
As Table \ref{TableI} shows, the transfer parameters only slightly decrease with the increase of Ln atomic number.
On the other hand, the gap $\Delta$ between the $4f$ and the $\pi^*$ orbital levels is obtained as huge as in the Gd complex
(Table \ref{TableI}), leading again to small promotion energy and, consequently, 
to the dominant role of the kinetic mechanism in the Ln$^{3+}$-N$_2^{3-}$ exchange coupling of complexes {\bf 2}-{\bf 5}. 
Given the small change of $t$ through {\bf 1}-{\bf 5}, 
the strong variation of the strength of exchange interaction  
in this series of complexes, testified by the experimental magnetic susceptibilities (Fig. \ref{Fig1}b), is expected to be due to the variation of the promotion energy.

\subsection*{Anisotropic exchange interaction}
Contrary to the Gd complex, the other members of the series are characterized by strong magnetic anisotropy 
on the Ln sites induced by the crystal-field (CF) splitting of their atomic $J$ multiplets.
These CF-split multiplets are described by multi-configurational wave-functions, therefore, they should be treated by explicitly correlated {\it ab initio} approaches \cite{Chibotaru2012, Ungur2015} rather than DFT.
The {\it ab initio} fragment calculations show that the CF 
split $J$ multiplet on Tb$^{3+}$ ion ($\approx 700$ cm$^{-1}$) 
is of the same order of magnitude as the estimated isotropic exchange splitting 
in {\bf 1} ($\approx 400$ cm$^{-1}$). 
Therefore, in sharp contrast with the common situation in lanthanides, the exchange coupling in the anisotropic {\bf 2}-{\bf 5} 
does not reduce to the interaction between individual (lowest) CF doublets on Ln sites with the $S=1/2$ spin of the radical, 
Eq. (\ref{Eq:H_Ising}), but will intermix the entire CF spectrum 
arising from the ground atomic $J$ multiplet at lanthanide ions. 
Then, such exchange interaction should be formulated 
in terms of the total angular momenta $\hat{\mathbf{J}}_i$ ($i=1,2$) on the lanthanide sites.

\begin{figure*}[tb]
 \begin{center}
  \includegraphics[width=14cm, bb=0 0 400 208]{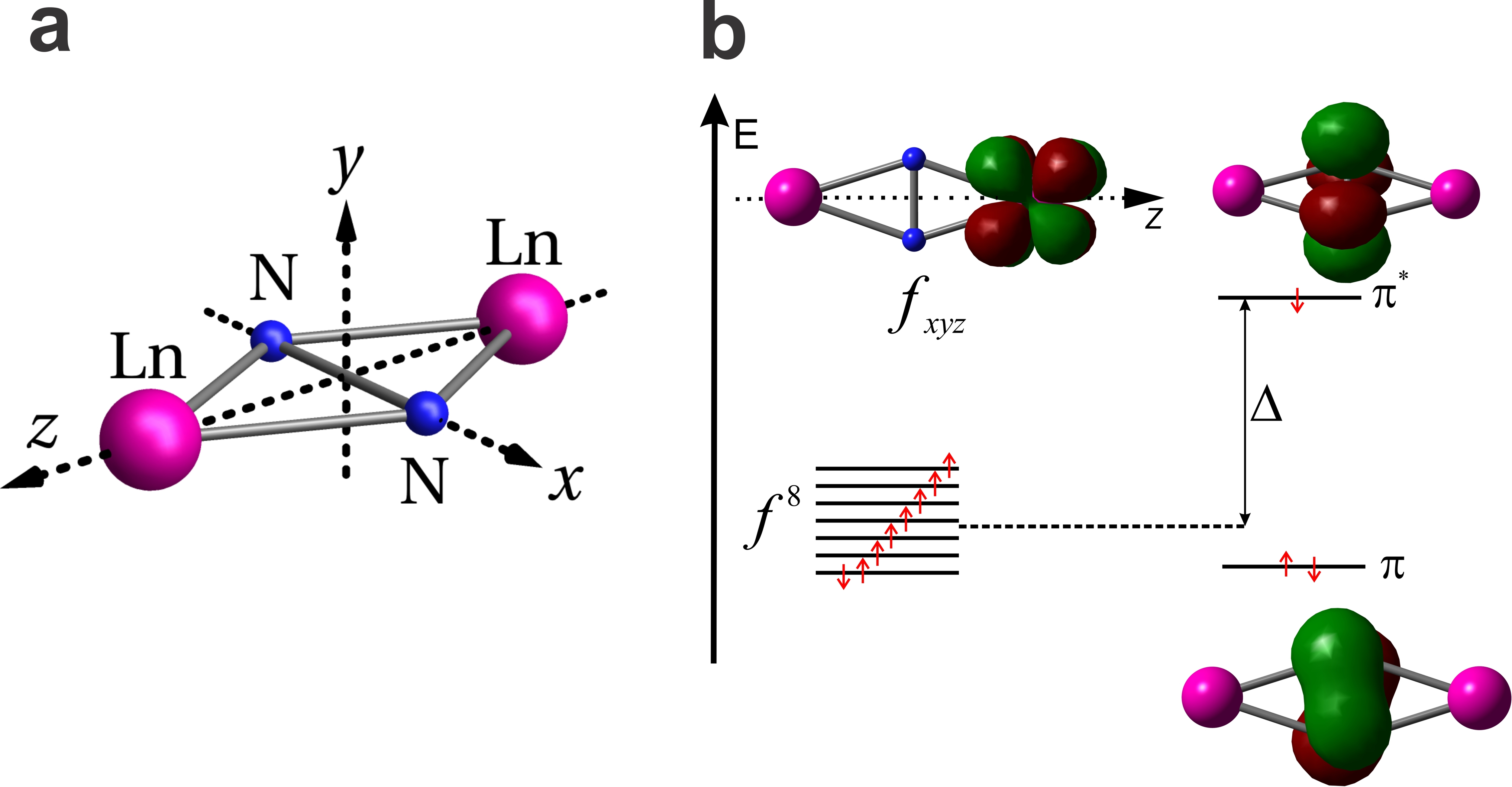}
 \end{center}
\caption{
{\bf a}, Exchange core Ln$^{3+}$-N$_2^{3-}$-Ln$^{3+}$ in the complex 
corresponding to $D_{2h}$ symmetry. 
{\bf b}, Magnetic orbitals in {\bf 1} obtained from DFT calculations. 
Only the $f$ orbital involved in the kinetic exchange mechanism is shown. 
}
\label{Fig2}
\end{figure*}

Extending the Anderson's superexchange theory \cite{Anderson1959, Anderson1963} to strong spin-orbit coupled systems, 
the tensorial form of the kinetic (covalent) 
interaction has been recently derived from the microscopic electronic Hamiltonian \cite{Iwahara2015}. 
The kinetic interaction 
between the lanthanide and radical centers 
contains besides the exchange part ($\hat{H}_\text{ex}$) also the Ln$^{3+}$-N$_2^{3-}$ covalent contribution 
(arising from Ln$^{3+}$-N$_2^{3-}$ electron delocalization) to the CF splitting at the Ln$^{3+}$ sites ($\hat{H}'_\text{cf}$):
\begin{eqnarray}
 \hat{H}'_\text{cf} &=& \sum_{i=1,2} 
                        \sum_{kq} 
                        \mathcal{J}_{kq00} 
                        \frac{O_k^q(\hat{\mathbf{J}}_i) \hat{I}}{O_k^0(J)},
\label{Eq:Hcov}
\\
 \hat{H}_\text{ex} &=& \sum_{i=1,2} 
                       \sum_{kqq'} 
                       \mathcal{J}_{kq1q'} 
                       \frac{O_k^q(\hat{\mathbf{J}}_i) \hat{S}_{q'}}{O_k^0(J)S}.
\label{Eq:Hex}
\end{eqnarray}
Here, 
$\hat{I}$ and $\hat{S}_q$ are the unit and the spin operators, respectively, of the radical's spin $S = 1/2$, 
$O_k^q(\hat{\mathbf{J}})$ are the Stevens operators \cite{Stevens1952} of rank $k$ and component $q$, 
and $\mathcal{J}_{kq00}$ and $\mathcal{J}_{kq1q'}$ are the exchange parameters 
\cite{Iwahara2015}.
The Stevens operator $O_k^q(\hat{\mathbf{J}})$ is a polynomial of $\hat{J}_\alpha$ $(\alpha=x,y,z)$ of $k$th degree, in which 
$|q|$ $(= \pm q)$ corresponds to the order of $\hat{J}_\pm$ $(= \hat{J}_x \pm i \hat{J}_y)$.
The maximal rank of $k$ is 7 for the considered Ln$^{3+}$ ions, whereas 
the maximal $|q|$ is 5 in the present case because only the $4f_{xyz}$ magnetic orbitals at the lanthanide sites contribute to the kinetic exchange. 
The summation over $k$ in Eqs. (\ref{Eq:Hcov}) and (\ref{Eq:Hex}) is confined to even and odd ranks, respectively,
which is required by the invariance of these Hamiltonians with respect to time-inversion. 
As it is seen from the form of these Hamiltonians, $\hat{H}'_\text{cf}$ only contributes to the CF splitting of $J$ multiplets on individual metal sites, 
whereas $\hat{H}_\text{ex}$ describes the interaction between powers of total angular momenta at the metal sites with components of spin $S=1/2$ of the N$_2^{3-}$ radical. 
For comparison, the weak anisotropic exchange interaction between two spins 
(pseudospins)
is described by the exchange Hamiltonian $\hat{\mathbf{S}}_1 \cdot \mathbf{D} \cdot \hat{\mathbf{S}}_2$, where $\mathbf{D}$ is the $3 \times 3$ exchange matrix, containing one isotropic, five symmetric anisotropic and three antisymmetric (Dzyaloshinsky-Moriya) exchange parameters \cite{Moriya1960}. 
This Hamiltonian corresponds to the first rank contribution ($k=1$) in Eq. (\ref{Eq:Hex}), where $\mathcal{J}_{1q1q'}$ are just the nine components of the above exchange matrix $\mathbf{D}$.
The expression for the exchange parameter $\mathcal{J}_{kqk'q'}$ \cite{Iwahara2015} includes all virtual electron transfer processes,
$(4f)^{n}(\pi^*)^1 \rightarrow (4f)^{n+1}(\pi^*)^0 \rightarrow (4f)^{n}(\pi^*)^1$, 
where $n$ is the number of $4f$ electrons in Ln$^{3+}$.
The multiplet electronic structure of Ln$^{2+}$ is fully included in 
the electron promotion energy $U_0 + \Delta E_\alpha$ and the wave functions of the intermediate states,
where by $U_0$ we further denote the smallest promotion energy,
$\alpha$ numbers the intermediate $J$-multiplets,
and 
$\Delta E_\alpha$ is the excitation energy of the multiplet $\alpha$ with respect to the ground one in Ln$^{2+}$.

The highly complex tensorial form of the exchange Hamiltonian 
is inevitable for orbitally degenerate systems 
with strong spin-orbit coupling, as was pointed out long time ago \cite{Elliott1968, Hartmann-Boutron1968}.
Although all exchange parameters $\mathcal{J}_{kqk'q'}$ are in principle required for adequate description 
of the exchange interaction, it is hardly possible to extract a sufficient large number of them from experiment in a unique way.
However, once $\mathcal{J}_{kqk'q'}$ are expressed via microscopic electronic parameters \cite{Iwahara2015},
the latter can be determined from up-to-date quantum chemistry calculations. Thus 
the transfer parameter $t$ is obtained here from DFT calculations, expected to be accurate enough \cite{dft1, dft2},
whereas the excitation energies $\Delta E_\alpha$ and the CF states 
are obtained by fragment state-of-the-art {\it ab initio} calculations including spin-orbit coupling \cite{Chibotaru2012, Ungur2015}.
The only parameter that might be inaccurate when extracted from DFT or {\it ab initio} calculations is $U_0$. Indeed,
the former gives at most an averaged value over multiplets $\bar{U}$ and 
the latter systematically overestimates it due to insufficient account of dynamical correlation.

\begin{table*}[tb]
\begin{ruledtabular}
\caption{
Transfer parameters $t$, 
energy gaps $\Delta$ between the $4f$ and the $\pi^*$ orbital levels,
minimal electron promotion energies $U_0$ (all in cm$^{-1}$), 
$g$-factors and angles between the magnetic moments on Ln$^{3+}$ and N$_2^{3-}$ ($\theta$) 
in the ground exchange KD, 
and  blocking barriers $E_\text{barrier}$ (cm$^{-1}$) 
for complexes 
{\bf 1}-{\bf 5}.
For $E_\text{barrier}$, both the experimental (exp.) \cite{Rinehart2011, Rinehart2011JACS} 
and present (calc.) data are shown. 
}
\label{TableI}
\begin{tabular}{cccccc} 
               & {\bf 1} (Gd) & {\bf 2} (Tb) & {\bf 3} (Dy) & {\bf 4} (Ho) & {\bf 5} (Er) \\
\hline
$t$            & 1407 & 1333 & 1322 & 1311 & 1270 \\
$\Delta$       & $5.20 \times 10^{4}$ & $5.74 \times 10^{4}$ & $5.80 \times 10^{4}$ & $5.73 \times 10^{4}$ & $5.78 \times 10^{4}$ \\
$U_0$          & 8500 & 4600 & 6500 & 7400 & 12200 \\
$g_x$    & $2.2 \times 10^{-2}$ & $7.6\times 10^{-6}$ & $2.2 \times 10^{-6}$ & $4.7 \times 10^{-3}$ & $1.3 \times 10^{-3}$ \\
$g_y$    & $3.7 \times 10^{-2}$ & $1.1\times 10^{-5}$ & $7.0 \times 10^{-6}$ & $1.2 \times 10^{-2}$ & $1.6 \times 10^{-3}$ \\
$g_z$    & 25.6                 & 33.6                & 37.5                 & 36.2                & 32.1 \\
$\theta$ & 0.0$^\circ$  & 2.5$^{\circ}$       & 2.3$^{\circ}$        & 2.6$^{\circ}$       & 6.2$^{\circ}$ \\
$E_\text{barrier}$ (exp.) & -    & 227  & 123  & 73   & 36   \\
$E_\text{barrier}$ (calc.)& -    & 208  & 121  & 105  & 28   \\
\end{tabular}
\end{ruledtabular}
\end{table*}

In this way we construct the full microscopic Hamiltonian, $\hat{H} = \hat{H}_\text{cf} + \hat{H}'_\text{cf} + \hat{H}_\text{ex}$, 
containing only
one unknown parameter $U_0$, 
where $\hat{H}_\text{cf}$ is the {\it ab initio} CF Hamiltonian for mononuclear Ln fragments (see Supplemental Materials for details). 
Diagonalizing this Hamiltonian, the magnetic susceptibility 
$\chi$ 
for the entire series of compounds has been simulated as described elsewhere \cite{Ungur2015}. 
Figure \ref{Fig1}b shows that the experiment is well reproduced for the values of minimal promotion energy 
$U_0$
listed in Table \ref{TableI}. 
The calculated exchange parameters for the 
series of the complexes are shown in Table \ref{TableII}.
We can see from the table that the exchange interaction 
involves non-negligible
contributions up to the rank $k=7$. 

\begin{figure*}[tb]
 \begin{center}
  \includegraphics[bb= 0 0 1166 892, width=16cm]{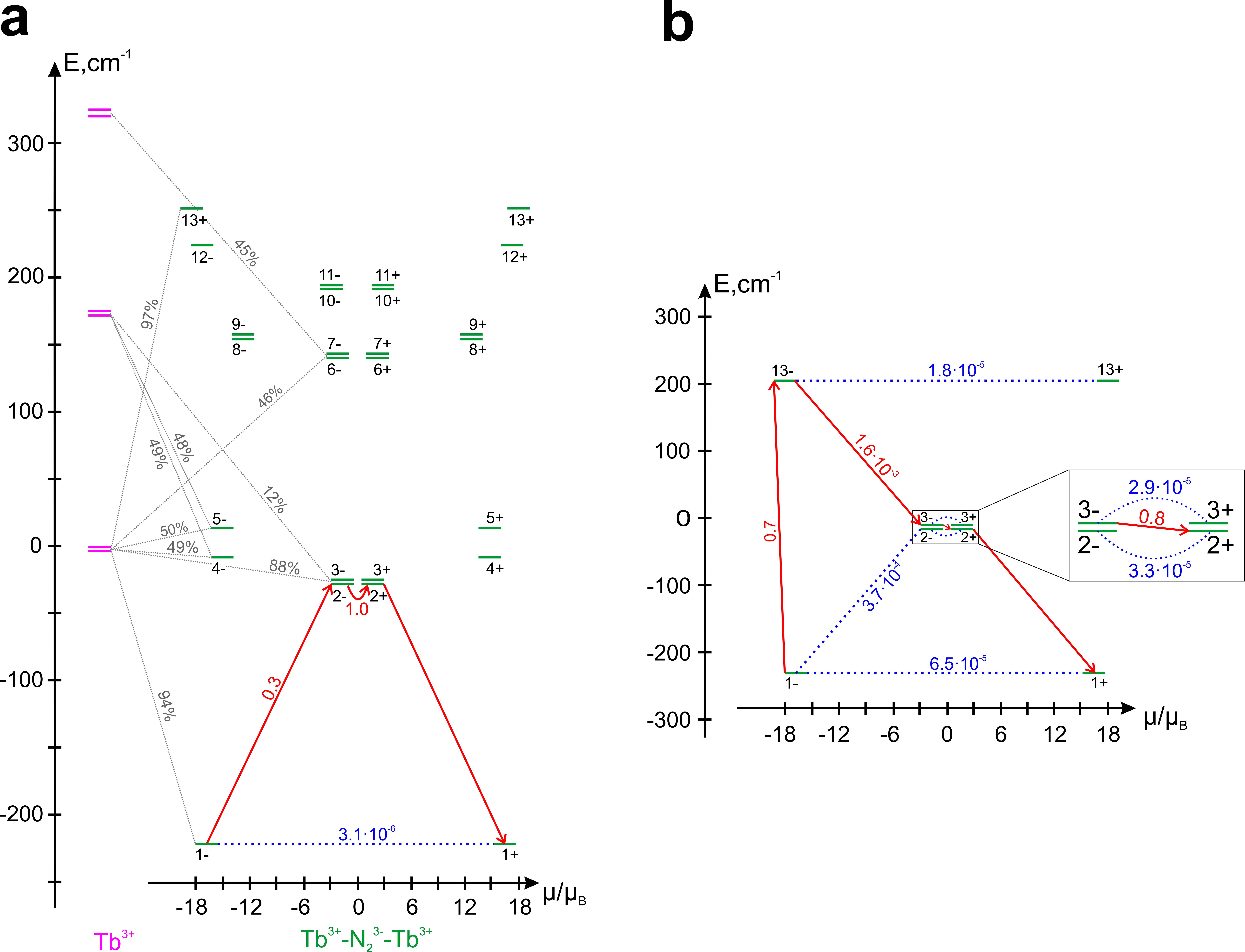}
 \end{center}
\caption{
 The low-lying exchange spectrum and the magnetization blocking barrier in {\bf 2}. 
{\bf a}, The violet bold lines show the CF levels on Tb ions,
the green bold lines show the low-lying exchange levels.  
Each exchange level is placed according to the projection of its 
magnetic moment on the main magnetic axis of the ground exchange doublet (green dashed line in Fig. \ref{Fig1}a). 
The exchange
levels with the same number are two components of the corresponding 
KD. 
The thin dashed lines show the admixed CF states on Tb sites to the exchange states 
in percent (only admixtures $>$ 10\% are shown).
The number accompanying the red line is the average magnetic moment matrix element (in $\mu_{\text{B}}$) between the components of the lowest exchange KD; 
the rate of QTM in the ground exchange state is proportional to its square.
The red arrows denote the relaxation path outlining the barrier of reversal of magnetization, with the same meaning of the corresponding numbers (see the text for more details). 
{\bf b}, The magnetization blocking barrier for {\bf 2} calculated in the absence of the admixture of excited CF states on Tb sites to the ground one via the exchange interaction. 
}
\label{Fig3}
\end{figure*}

The low-lying exchange spectrum 
for the Tb complex is shown in Fig. \ref{Fig3}a. 
The ground 
($1\pm$)
and the first two excited 
($2\pm, 3\pm$)
exchange Kramers doublets (KDs) mainly originate from the ground CF doublets on the Tb ions 
(94 \%, 87 \%, and 88 \%, respectively). 
However, the third and fourth excited exchange KDs 
($4\pm, 5\pm$)
represent almost equal mixtures of the ground and the first excited 
CF doublets on the Tb$^{3+}$ sites. This is remarkable because the mixed CF states are separated by 166 cm$^{-1}$ 
(Fig. \ref{Fig3}a). 
Similar scenario is realized in {\bf 3} and {\bf 4}, whereas in {\bf 5}
the exchange interaction and the resulting mixing of CF states is relatively weak.
The magnetic structure of the ground exchange KD is shown in Fig. \ref{Fig1}a.
The magnetic moments on Tb$^{3+}$ sites are parallel due to inversion symmetry and almost coincide with
the directions of the main magnetic axes in the ground local KDs (Fig. \ref{Fig1}a). 
The magnetic moment of the radical, corresponding to isotropic $S = 1/2$, is rotated 
with respect to the magnetic moments on Tb sites by small angle 
$\theta$ (Table \ref{TableI})
due to the non-Heisenberg contributions to the exchange interaction \cite{Chibotaru2015Ising}.

\begin{table}[tb]
\begin{ruledtabular}
\caption{
Calculated exchange parameters $\mathcal{J}_{kq1q'}$ (cm$^{-1}$) for the complexes {\bf 1}- {\bf 5}.
}
\label{TableII}
\begin{tabular}{cccccccc}
$k$ & $q$  & $q'$   & \multicolumn{5}{c}{$\mathcal{J}_{kq1q'}$}\\
    &      &        & {\bf 1} (Gd) & {\bf 2} (Tb) & {\bf 3} (Dy) & {\bf 4} (Ho) & {\bf 5} (Er) \\
\hline
1 & 0      & 0      &   94.9 &   95.8   &   70.8  &   55.4  &   24.2  \\
1 & $\pm$1 & $\mp$1 & $-94.9$& $-95.8$  & $-70.8$ & $-55.4$ & $-24.2$ \\
3 & 0      & 0      & $0.0$  &   13.4   & $-10.6$ & $-4.4$  &    5.0  \\
3 & $\pm$1 & $\mp$1 & $0.0$  &    8.2   & $-6.5$  & $-2.7$  &    3.0  \\
3 & $\pm$3 & $\pm$1 & $0.0$  &   10.6   & $-8.4$  & $-3.5$  &    3.9  \\
5 & 0      & 0      & $0.0$  &   17.0   & $-16.0$ & $-1.6$  &    4.2  \\
5 & $\pm$1 & $\mp$1 & $0.0$  & $-12.8$  &   8.4   &   6.8   &  $-6.1$ \\
5 & $\pm$3 & $\pm$1 & $0.0$  & $-2.5$   &   7.5   & $-7.6$  &    3.4  \\
5 & $\pm$4 & 0      & $0.0$  &   5.7    & $-0.8$  & $-7.5$  &    5.0  \\
5 & $\pm$5 & $\mp$1 & $0.0$  & $-13.5$  &   11.5  &   3.2   &  $-4.4$ \\
7 & 0      & 0      & $0.0$  &   0.3    & $-3.3$  &   4.6   &  $-2.3$ \\
7 & $\pm$1 & $\mp$1 & $0.0$  & $-0.2$   &   2.5   & $-3.5$  &    1.7  \\
7 & $\pm$3 & $\pm$1 & $0.0$  &   0.2    & $-2.2$  &   3.0   &  $-1.5$ \\
7 & $\pm$4 & 0      & $0.0$  & $-0.4$   &   5.1   & $-7.1$  &    3.5  \\
7 & $\pm$5 & $\mp$1 & $0.0$  &   0.6    & $-7.2$  &  10.0   &  $-5.0$ \\
\end{tabular}
\end{ruledtabular}
\end{table}

One may notice that the dominant first rank term of the exchange interaction 
is of isotropic Heisenberg type despite the strong spin-orbit coupling in Ln$^{3+}$ ions 
(Table \ref{TableII}).
This looks surprising because even weak spin-orbit coupling
makes the first-rank exchange interaction anisotropic \cite{Moriya1960}.
The analysis of the expression for the first-rank exchange parameters $\mathcal{J}_{1q1q'}$ \cite{Iwahara2015} shows that they are in general of non-Heisenberg type, 
whereas the present case is the only possible exception (see Supplemental Material). 
Indeed, the isotropy of the first-rank exchange contribution requires involvement of only $f$ orbitals with the 
projections $m=\pm 2$.
This can only arise for high symmetry of the exchange bridge 
(Fig. \ref{Fig2}a)
and for situations with one single electron transfer path, as in the present case.
If any other orbital (or more of them) contribute to the electron transfer, 
the first-rank exchange interaction becomes strongly anisotropic.

\subsection*{Magnetization blocking barriers}
Table \ref{TableI} shows that the transverse $g$-factors ($g_x$ and $g_y$) in the ground exchange KD,
the squares of which characterize the rate of quantum tunneling of magnetization (QTM) \cite{Chibotaru2015}, 
are the largest for {\bf 4} and the smallest for {\bf 2} and {\bf 3} complexes. 
This explains why large magnetization hysteresis is seen at low temperatures in the latter two compounds, 
while not seen at all in the former and only weakly observed in the complex {\bf 5} \cite{Rinehart2011,Rinehart2011JACS}. 
The path characterizing the activated magnetic relaxation in high-temperature domain is shown for the Tb complex 
in Fig. \ref{Fig3}a by blue arrows. 
The height of the activation barrier 
$E_\text{barrier}$
corresponds to the first excited exchange KD, because its two components 
($2\pm$ in Fig. \ref{Fig3}a) are connected by a large magnetic moment matrix element which causes a large 
temperature-assisted QTM. 
Blocking barriers of similar structure 
(Fig. \ref{Fig3}a)
arise in {\bf 3} and {\bf 4}, their calculated activation energies comparing well with the experimental ones (Table \ref{TableI}). 

The unusually large matrix elements between the ground and the first excited exchange KDs are entirely 
due to the exchange mixing of the ground and the first excited CF doublets on the Ln sites. 
Indeed, if one quenches the exchange admixture of excited CF doublets to the ground ones, this matrix element 
becomes three orders of magnitude smaller (Fig. \ref{Fig3}b). 
Then the activated relaxation will proceed via a higher exchange doublet, thereby doubling the height of the 
blocking barriers (Fig. \ref{Fig3}b). 
Thus in the case of very strong exchange interaction, which is able to intermix the CF states on Ln sites, 
the axiality of the ground and excited exchange doublets is diminished dramatically and the blocking barriers 
do not exceed the energy of the first excited exchange KD. 
In other words, the strength of exchange interaction after reaching a certain value starts playing a 
destructive role for the magnetization blocking. 
Therefore, to exploit the effect of strong exchange interaction for achieving high magnetization blocking, 
an even stronger axial CF field on the Ln sites, precluding the exchange admixture of excited CF states, 
seems to be indispensable.

\section*{Discussion}
The mixed lanthanide complexes {\bf 1}-{\bf 5} investigated in this work are unique because they show an exchange interaction up to two orders of magnitude stronger than in conventional lanthanide complexes. Due to such strong exchange interaction, a qualitatively new situation arises when the exchange coupling starts to intermix the CF multiplets on the Ln sites. In all previous lanthanide complexes only the ground CF doublets on Ln sites were involved, which led to conventional Ising-type exchange interactions. In the present case, due to the involvement of all CF doublets belonging to the atomic $J$-multiplet, the exchange interaction becomes highly complex, requiring a tensorial description and involving many parameters.

By combining DFT and {\it ab initio} calculations with
the microscopic modeling of the exchange interaction, we were able to unravel
the mechanism of giant exchange interaction in these complexes. 
This exchange interaction is found to be kinetic and highly complex, involving non-negligible contributions up to seventh power of total angular
momentum of each Ln site. 
Based on the calculated exchange states, the mechanism of the magnetization blocking is revealed. 
Contrary to general expectations the latter is not always favored by strong exchange interaction.
The accuracy of our approach is proved by the close reproduction of experimental
magnetic susceptibility and magnetization blocking barrier for all investigated compounds. 

The theoretical analysis proposed in this work opens the way for the investigation of highly complex exchange interaction in materials with strongly anisotropic magnetic sites. Given the large number of involved exchange parameters and the obvious difficulties of their experimental determination, such an approach can become a powerful tool for the study of magnetic materials of primary interest.

\section*{Methods}
\subsection*{DFT calculations.}
All DFT calculations were carried out with ORCA 3.0.0. program \cite{orca} 
using the B3LYP functional and SVP basis set. 
Scalar relativistic effects were taken into account within Douglas-Kroll-Hess Hamiltonian. 
The isotropic exchange parameter for the complex {\bf 1}, $\mathcal{J}_\text{Heis}$,
was derived by applying the broken-symmetry approach \cite{Soda2000}.
The obtained $\mathcal{J}_\text{Heis}$ was divided by 2
to account for its overestimation due to the self-interaction error \cite{Polo2003, Ruiz2005}. 
The $4f$ and the $\pi^*$ orbital levels and the transfer parameters $t$ for all complexes {\bf 1}-{\bf 5}
were derived by projecting the Kohn-Sham orbitals onto a tight-binding model. 
The averaged promotion energy $\bar{U}$ for the complex {\bf 1} was derived 
by reproducing the energy difference between the high-spin and the broken-symmetry DFT states with a Hubbard model. 

\subsection*{{\it Ab initio} calculations.} 
Energies and wave functions of CF multiplets on Ln$^{3+}$ sites in {\bf 1}-{\bf 5} have been obtained from fragment 
{\it ab initio} calculations including the spin-orbit coupling, using the quantum chemistry package Molcas 7.8 \cite{molcas}. 
The calculations have been done for the experimental geometry of the complexes, 
in which one of the two Ln$^{3+}$ ions was replaced by an isovalent closed-shell La$^{3+}$ ion. 
The total number of electrons was reduced by unity in order to have a closed-shell electronic configuration N$_2^{2-}$ 
on the dinitrogen bridge. 
To simulate the electrostatic crystal field from the removed radical's electron, 
two point charges of $-0.5 e$ were added on the nitrogen atoms. 
For this structural model of a Ln fragment, the complete active space self-consistent field (CASSCF) approach 
was used including all seven $4f$ orbitals of the Ln atom in the active space. 
The spin-orbit interaction was treated with the module SO-RASSI and the local magnetic properties 
were calculated with the SINGLE\_ANISO module of Molcas \cite{single_aniso}.
Exchange energy spectrum and magnetic properties of the investigated polynuclear compounds were calculated using the 
POLY\_ANISO program \cite{Ungur2015, single_aniso}, modified to treat the general form of exchange interaction, 
Eqs. (2), (3), within the kinetic exchange mechanism.

For further details, see Supplemental Material.


%

\clearpage
\begin{center}
\textbf{
Supplemental Materials\\
for
\\
``Giant exchange interaction in mixed lanthanides''
}
\end{center}
\setcounter{equation}{0}
\setcounter{figure}{0}
\setcounter{table}{0}
\setcounter{section}{0}
\makeatletter
\renewcommand{\theequation}{S\arabic{equation}}
\renewcommand{\thefigure}{S\arabic{figure}}
\renewcommand{\thetable}{S\arabic{table}}
\renewcommand{\bibnumfmt}[1]{[S#1]}
\renewcommand{\citenumfont}[1]{S#1}

This material contains: 
\\
1) DFT based derivations of the $4f$ and the $\pi^*$ orbital levels and of the transfer parameters 
$t$ for all complexes {\bf 1}-{\bf 5};
\\
2) Fragments {\em ab initio} calculations of the energies and wave functions of CF multiplets on Ln$^{3+}$ sites in {\bf 1}-{\bf 5}, and calculations of atomic multiplets of the corresponding Ln$^{2+}$ ions; 
\\
3) The calculation of the exchange spectra are described;
\\
4) The analysis of the first rank exchange parameters.


\section{DFT calculations} 

\subsection{Extraction of the transfer parameter $t$ for {\bf 1}-{\bf 5}}
In order to derive the transfer parameters between the $4f$ orbital and 
the $\pi^*$ orbital of the bridging N$_2$,
the Kohn-Sham levels are projected into tight-binding Hamiltonian: 
\begin{eqnarray}
 \hat{H} &=& \sum_\sigma \left[\sum_{i=1}^2 \epsilon_f \hat{n}_{i\tilde{\gamma}\sigma} 
           + \epsilon_{\pi^*} \hat{n}_{\pi^*\sigma}
\right.
\nonumber\\
          & +& 
\left.
 t \left(
               \hat{c}_{1\tilde{\gamma}\sigma}^\dagger \hat{c}_{\pi^*\sigma} +
               \hat{c}_{\pi^*\sigma}^\dagger \hat{c}_{1\tilde{\gamma}\sigma} -
               \hat{c}_{2\tilde{\gamma}\sigma}^\dagger \hat{c}_{\pi^*\sigma} -
               \hat{c}_{\pi^*\sigma}^\dagger \hat{c}_{2\tilde{\gamma}\sigma} 
               \right)
           \right],
\nonumber\\
\label{Eq:HTBSM}
\end{eqnarray}
where $i (= 1,2)$ is the index for the Ln$^{3+}$ site in the complex,
N$_2^{3-}$ site is described by the type of the magnetic orbital $\pi^*$,
$\tilde{\gamma}$ is the orbital component $xyz$, 
$\sigma = \uparrow, \downarrow$ is the projection of spin operator, 
$\epsilon_f$ and $\epsilon_{\pi^*} (= \epsilon_f + \Delta)$ are one electron orbital levels 
of the $4f$ orbital and the $\pi^*$ orbital, respectively,
$t$ is the transfer paremeter between the $4f$ and the $\pi^*$ orbitals,
$\hat{c}^\dagger$ ($\hat{c}$) is an electron creation (annihilation) operator, 
and $\hat{n}$ is a number operator.
The subscripts of the creation, annihilation, and number operators indicate 
the site, the orbital index for only lanthanide site, and spin projection.
Because of the $D_{2h}$ symmetry of the magnetic core part, only one $4f$ orbital ($4f_{xyz}$)
overlaps with the $\pi^*$ orbital
(Fig. 2b in the main text).
Therefore, we only include the $4f_{xyz}$ orbital for each lanthanide site in the model Hamiltonian.

Diagonalizing the tight-binding Hamiltonian (\ref{Eq:HTBSM}), the one-electron levels are obtained as 
\begin{eqnarray}
 \epsilon_{f,a} &=& \epsilon_f, 
\label{Eq:e_faSM}
\\
 \epsilon_{f,s} &=& \epsilon_f + \frac{1}{2}\left(\Delta - \sqrt{\Delta^2 + 8t^2}\right), 
\label{Eq:e_fsSM}
\\
 \epsilon_{\pi^*} &=& \epsilon_f + \frac{1}{2}\left(\Delta + \sqrt{\Delta^2 + 8t^2}\right),
\label{Eq:e_piSM}
\end{eqnarray}
where the subscript ``$a$'' and ``$s$'' indicate antisymmetric and symmetric orbitals, respectively.
Comparing these orbital levels with the DFT calculations, we obtain parameters $\epsilon_f$, $t$, and $\Delta$.

\begin{figure*}[htb]
\begin{center}
\begin{tabular}{lllll}
(a) & ~~~& (b) & ~~~& (c)
\\
\includegraphics[bb=0 0 2594 1775, height=4cm]{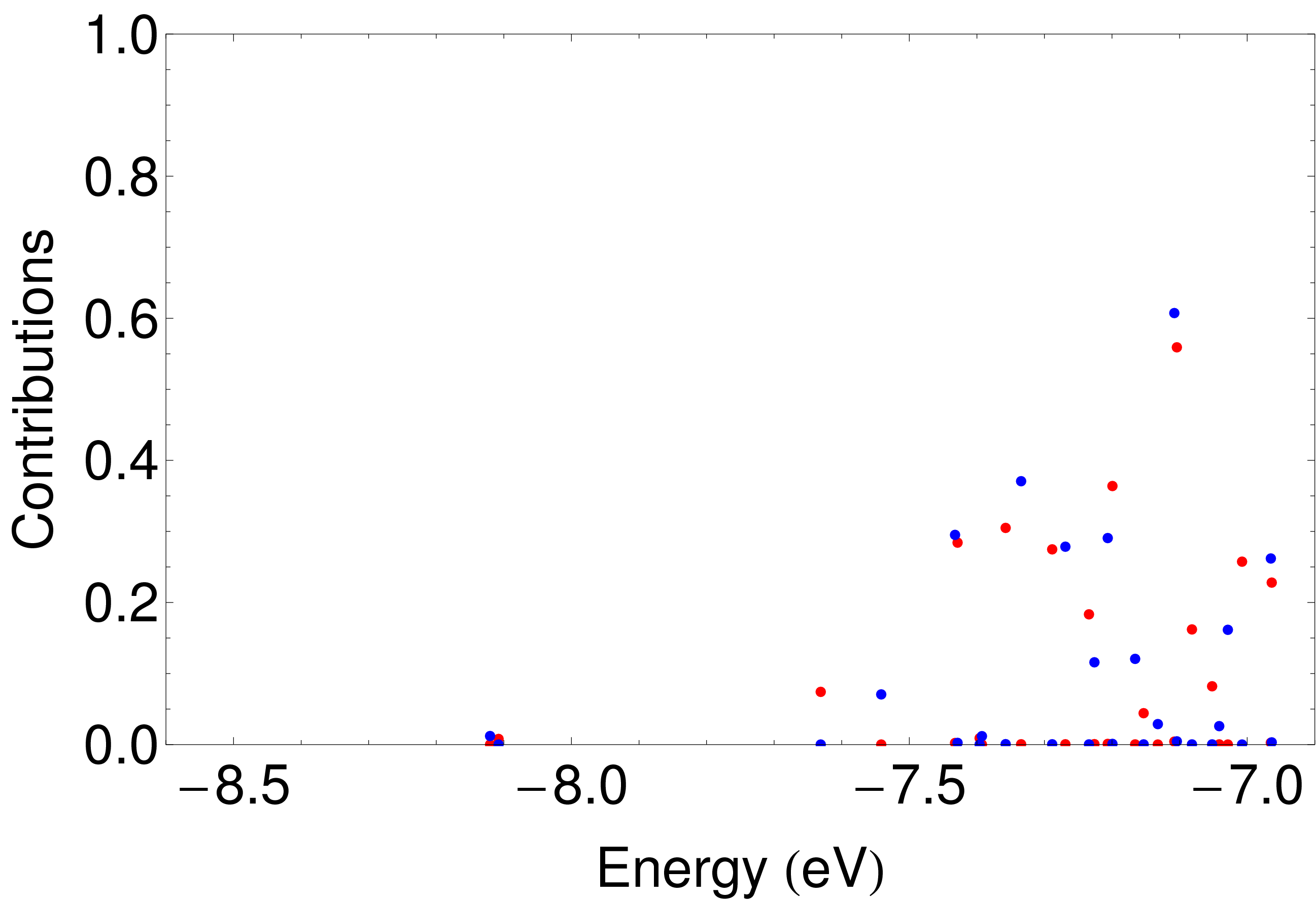}
& ~~~&
\includegraphics[bb=0 0 2594 1775, height=4cm]{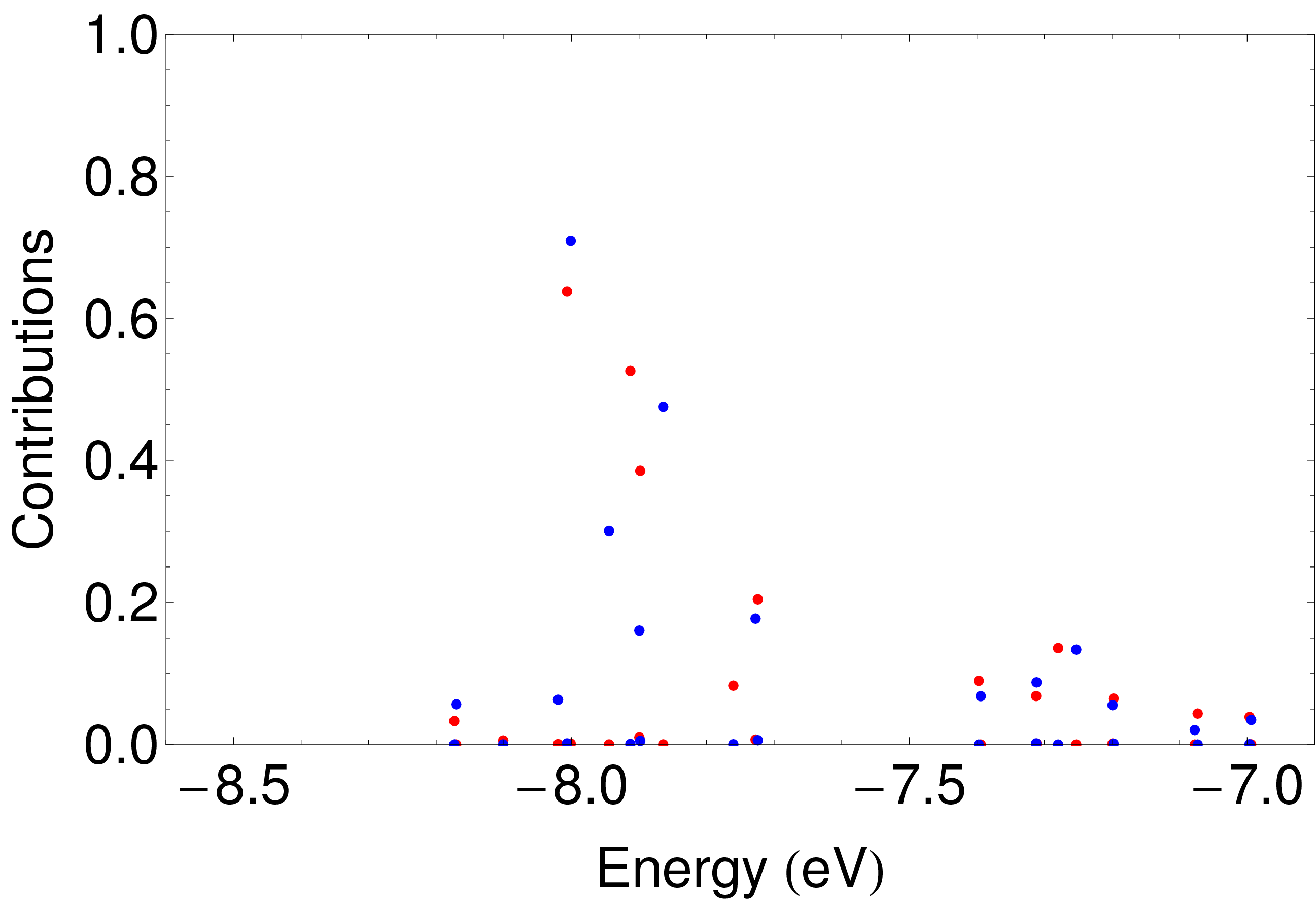}
& ~~~&
\includegraphics[bb=0 0 2594 1775, height=4cm]{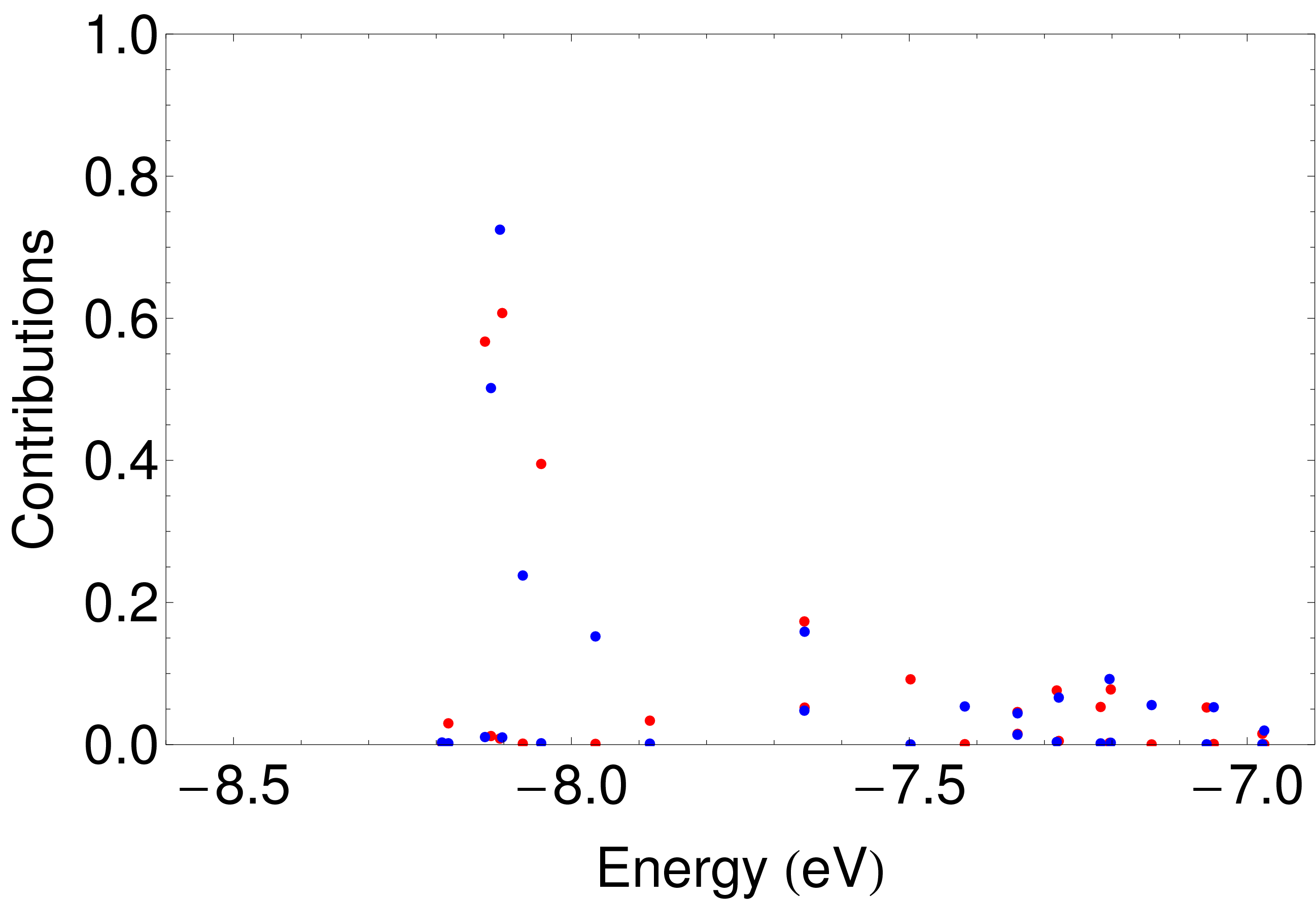}
\\
(d) & ~~~& (e)
\\
\includegraphics[bb=0 0 2594 1775, height=4cm]{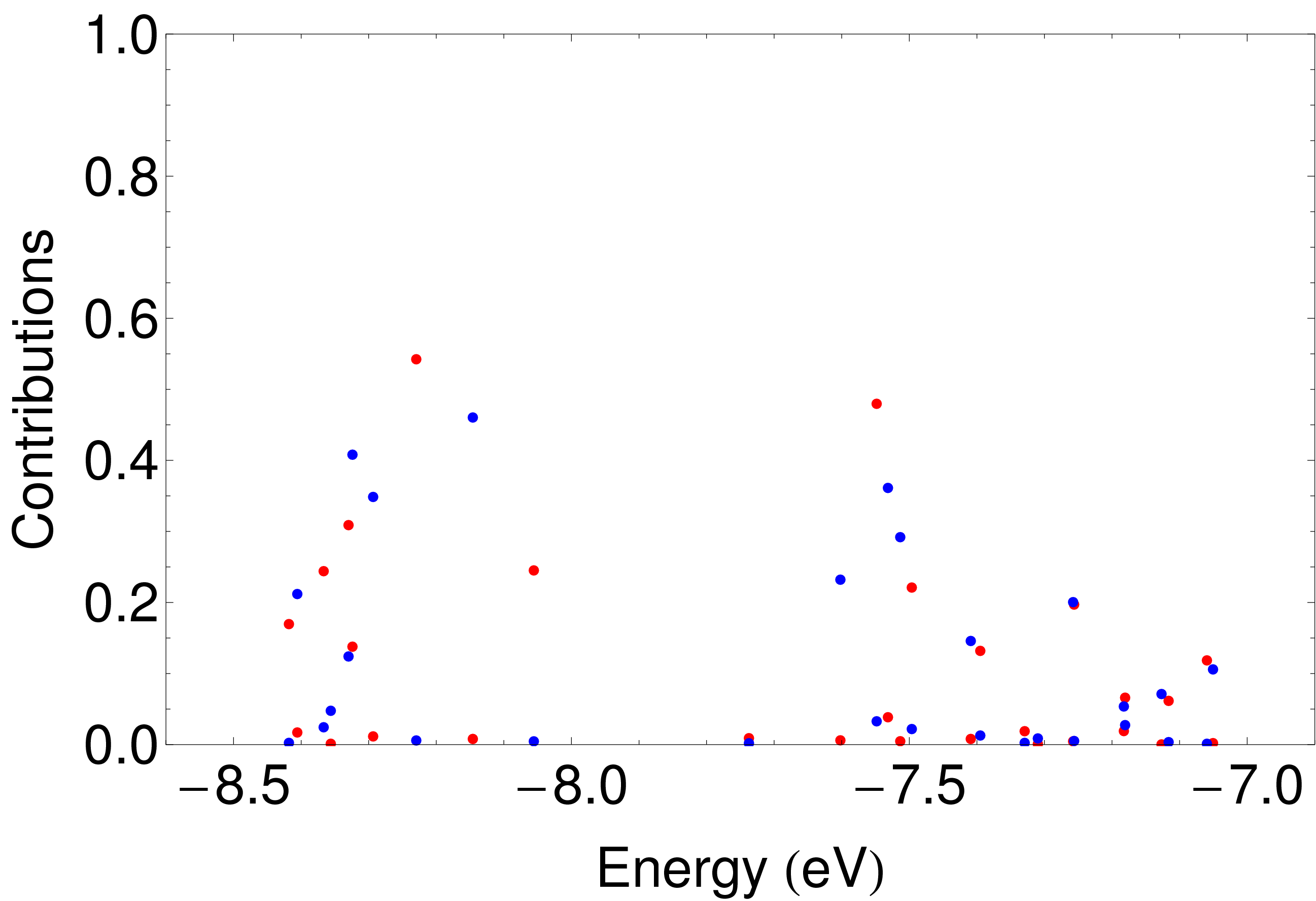}
& ~~~&
\includegraphics[bb=0 0 2594 1775, height=4cm]{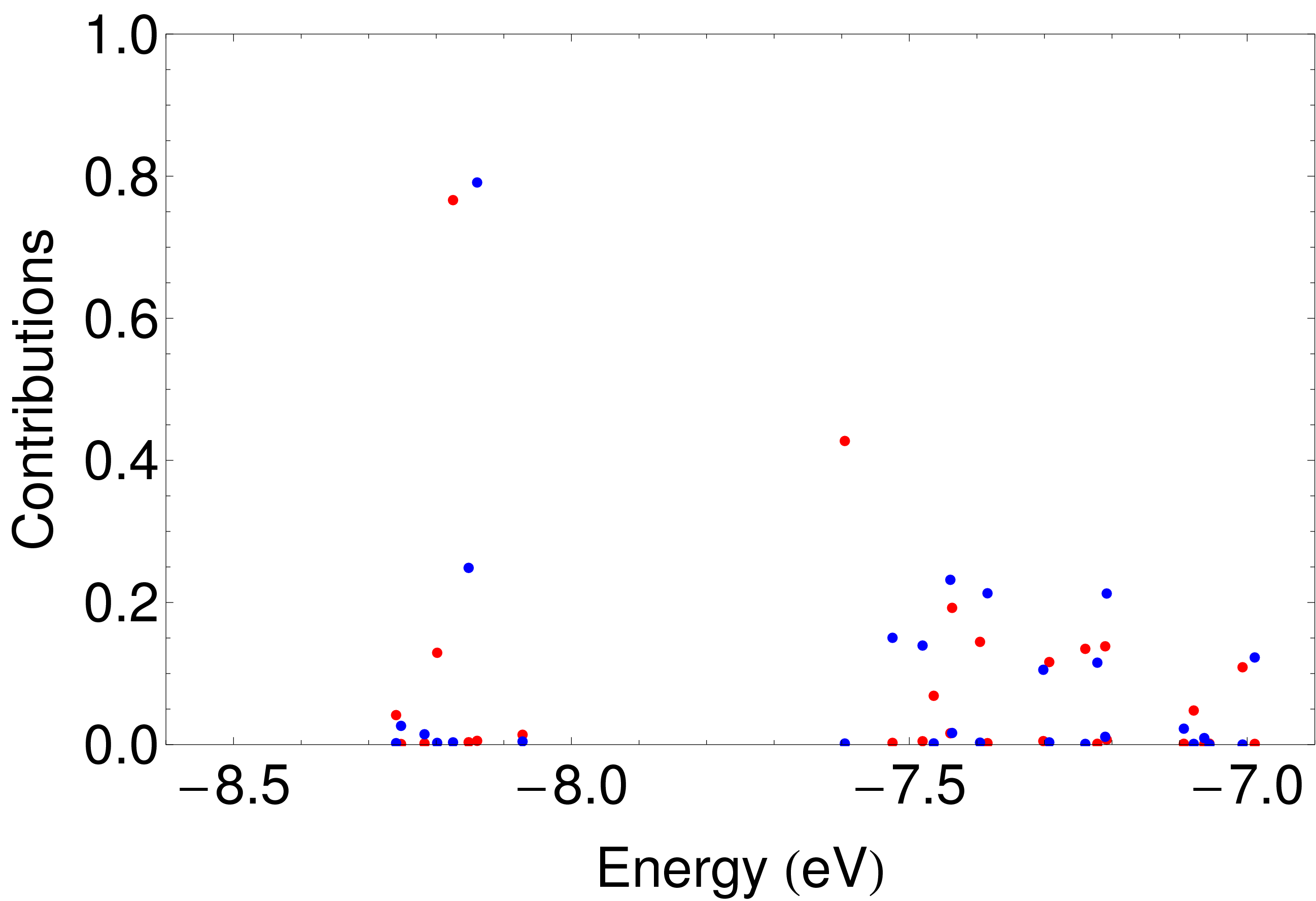}
\end{tabular}
\end{center}
\caption{Contributions of the antisymmetric $|C_{a,i}|$ (red) 
and the symmetric $|C_{s,i}|$ (blue) combinations of the $4f_{xyz}$ 
orbitals to each Kohn-Sham orbitals for the (a) Gd, (b) Tb, (c) Dy, (d) Ho, (e) Er complexes.}
\label{Fig:KS}
\end{figure*}

The highest occupied Kohn-Sham orbital for the down spin in the low-symmetry DFT solutions 
correspond to the $\pi^*$ orbital.
On the other hand, $4f$ atomic orbitals contribute to many Kohn-Sham orbitals.
Thus, the $4f$ orbitals are localized as follows.
Because of the inversion symmetry of the complexes, 
the $4f$ orbital part of each Kohn-Sham orbital $\psi_i$ is 
decomposed into the antisymmetric and symmetric parts:
\begin{equation}
 |\psi_i\rangle = 
 \frac{1}{\sqrt{2}}\left(|1\rangle + |2\rangle \right)C_{a,i} +
 \frac{1}{\sqrt{2}}\left(|1\rangle - |2\rangle \right)C_{s,i},
\label{Eq:CSM}
\end{equation}
where, $|1\rangle$ and $|2\rangle$ indicate the $4f_{xyz}$ orbitals on the first and the second 
lanthanide sites, respectively.
The absolute values of $C_{a,i}$ and $C_{s,i}$ for the occupied Kohn-Sham orbitals for the up spin part 
are shown in Fig. \ref{Fig:KS}.
As the antisymmetric and the symmetric levels, we averaged the Kohn-Sham levels: 
\begin{equation}
 \epsilon_{f,a} = \frac{\sum_{i}^\text{occ.} C_{a,i}^2 \epsilon_i}{\sum_{i}^\text{occ.} C_{a,i}^2},
 \qquad
 \epsilon_{f,s} = \frac{\sum_{i}^\text{occ.} C_{s,i}^2 \epsilon_i}{\sum_{i}^\text{occ.} C_{s,i}^2}.
\label{Eq:KS_fSM}
\end{equation}
In Eq. (\ref{Eq:KS_fSM}), the sum is taken over occupied Kohn-Sham orbitals.
With the use of the levels, the parameters $t$ and $\Delta$ are derived (Table I in the main text). 
The transfer parameter is gradually decreasing as the increase of the atomic number
because the ionic radius of the lanthanide shrinks.

\subsection{Calculation of $\pi^* \rightarrow 4f$ electron promotion energy for {\bf 1}}
The high- and low-spin states of the complex {\bf 1} were analyzed based on the Hubbard Hamiltonian:
\begin{eqnarray}
 \hat{H} &=& 
  \sum_{i=1,2} \sum_{\gamma \sigma} \epsilon_f \hat{n}_{i\gamma \sigma}
+ \sum_\sigma \epsilon_{\pi^*} \hat{n}_{\pi^* \sigma}
\nonumber\\
 &+& \sum_\sigma t \left(
     \hat{c}_{1\tilde{\gamma}\sigma}^\dagger \hat{c}_{\pi^*\sigma} +
     \hat{c}_{\pi^*\sigma}^\dagger \hat{c}_{1\tilde{\gamma}\sigma} -
     \hat{c}_{2\tilde{\gamma}\sigma}^\dagger \hat{c}_{\pi^*\sigma} -
     \hat{c}_{\pi^*\sigma}^\dagger \hat{c}_{2\tilde{\gamma}\sigma} 
     \right)
\nonumber\\
 &+& \sum_{i=1,2} \sum_{\langle \gamma \sigma, \gamma'\sigma'\rangle}
 u_f \hat{n}_{i\gamma \sigma} \hat{n}_{i\gamma' \sigma'}
 + u_{\pi^*} \hat{n}_{\pi^*\uparrow} \hat{n}_{\pi^*\downarrow} 
\nonumber\\
 &+&
   \sum_{i=1,2} \sum_{\gamma \sigma} \sum_{\sigma'} 
 v \hat{n}_{\gamma \sigma} \hat{n}_{\pi^*\sigma'},
\label{Eq:HubbardHSM}
\end{eqnarray}
where $\gamma$ is the component of the $4f$ orbital,
$u_f$ and $u_{\pi^*}$ are the intrasite Coulomb repulsions on Gd and N$_2$ sites, respectively, 
and $v$ is the intersite Coulomb repulsion between the Gd and N$_2$ sites.

The high-spin state with the maximal projection is described by one electron configuration:
\begin{equation}
 |1\uparrow, \pi^*\uparrow, 2\uparrow\rangle,
\end{equation}
where 1 and 2 are the lanthanide sites and $\uparrow$ and $\downarrow$ are spin projections. 
The $4f$ electrons which are not in the $4f_{xyz}$ orbital are not explicitly written here.
The total energy $E_{\rm HS}$ is 
\begin{equation}
 E_{\rm HS} = E_0 + (2n+1) \epsilon_f + \Delta + 2n v + n(n-1) u_f,
\label{Eq:EHSSM}
\end{equation}
where $E_0$ is the total electronic energy except for the electrons 
in the $4f$ orbitals and $\pi^*$ orbitals,
and $n$ is the number of the $4f$ electrons in Gd$^{3+}$ ion.  
For the low-spin state ($\uparrow,\downarrow,\uparrow$ type), the basis set is
\begin{equation}
 \{
 |1\uparrow, 1\downarrow, 2\uparrow\rangle, 
 |1\uparrow, \pi^*\downarrow, 2\uparrow\rangle, 
 |1\uparrow, 2\downarrow, 2\uparrow\rangle
 \}.
\end{equation}
Here, the configurations with the electron transfer from the $4f$ to the $\pi^*$ are not included
because these configurations do not contribute much to the low-energy states due to the 
large energy gap $\Delta$ between the $4f$ and the $\pi^*$ levels.
The lowest energy is 
\begin{eqnarray}
 E_{\rm LS} &=& E_0 + (2n+1)\epsilon_f + n(n-1)u_f 
\nonumber\\
 &+& 
 \frac{1}{2} \left(
  \Delta + 2nv + nu_f - \sqrt{\left(\Delta + 2nv - nu_f\right)^2 + 8t^2}
 \right).
\nonumber\\
\label{Eq:ELS}
\end{eqnarray}
The energy difference between the low- and high-spin states are 
\begin{eqnarray}
 \Delta E &=& E_{\rm LS} - E_{\rm HS} \\
 &=& \frac{1}{2} \left[
  nu_f - \left(\Delta + 2nv\right) - \sqrt{\left(\Delta + 2nv - nu_f\right)^2 + 8t^2}
 \right]
\nonumber\\
\\
 &=& \frac{1}{2} \left(
  \bar{U} - \sqrt{\bar{U}^2 + 8t^2}
 \right),
\label{Eq:DeltaE}
\end{eqnarray}
where 
\begin{eqnarray}
\bar{U} = nu_f - \Delta - 2nv
\label{Eq:U}
\end{eqnarray}
is the (averaged) electron promotion energy.
Eq. (\ref{Eq:U}) shows that 
(i) the energy gap $\Delta$ significantly reduces the promotion energy
and (ii) the promotion energy increases with the number of the $4f$ electrons $n$. 
Using the transfer parameter $t$ derived from the Kohn-Sham orbital, 
energy gaps between the high-spin state and low-spin state, and Eq. (\ref{Eq:DeltaE}), 
the averaged promotion energy $\bar{U}$ is derived. 

\section{{\it Ab initio} calculations}
\subsection{Fragment calculations for Ln$^{3+}$ centers in {\bf 1}-{\bf 5}}
\label{Sec:fragment}
\begin{figure}[tb]
 \begin{center}
  \includegraphics[width=7cm, bb=0 0 3339 2196]{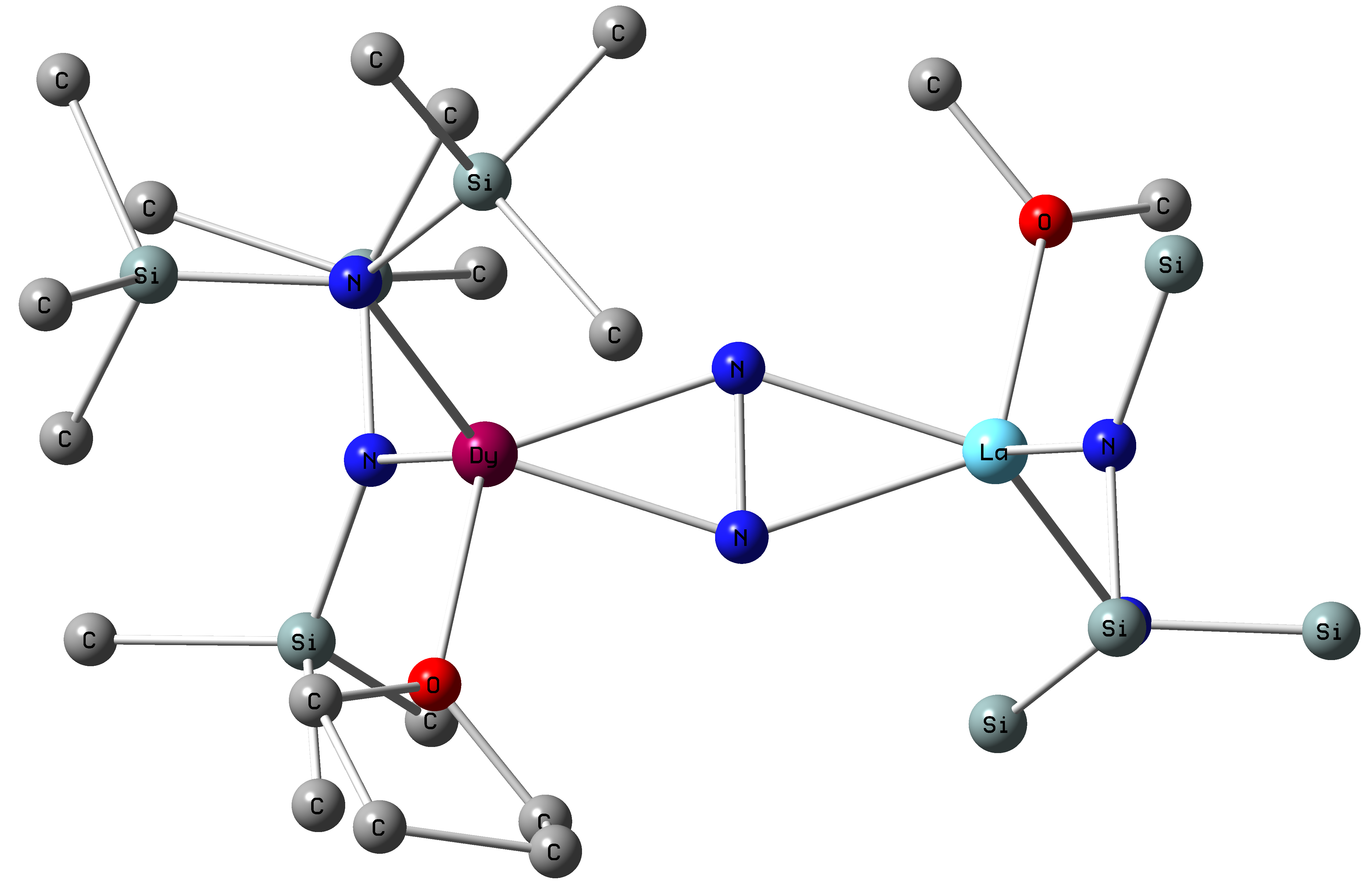}
 \end{center}
\caption{The LnLaN$_2^{3-}$ fragment used in {\it ab initio} calculations. 
Hydrogen atoms are omitted for clarity.
The right lanthanide ion was replaced by La in the {\it ab initio} calculations.}
\label{LnLaN2SM}
\end{figure}

To obtain the local electronic properties of the magnetic ions, 
{\it ab initio} quantum chemistry calculations (CASSCF/SO-RASSI) were performed using Molcas \cite{molcasSM}. 
In the calculations, one of the metal ions in the complex was replaced 
by diamagnetic lanthanum ion (La$^{3+}$) and the ligands for the La ion were reduced (Fig. \ref{LnLaN2SM}).
Two point charges ($-0.5$ $e$) were put on each N atom creating the N$_2$ bridge,
where $e$ $(>0)$ is the elementary charge.
The latter is to include the electrostatic potential from the unpaired electron of N$_2^{3-}$ bridge.
The covalent effect is included later ($\hat{H}'_\text{cf}$ in the main text).
In the CASSCF calculations, all $4f$ orbitals of the magnetic site are included in the active orbitals. 
The spin-orbit coupling is included in the SO-RASSI calculation. 
In the SO-RASSI calculations the following CASSCF states were mixed by spin-orbit coupling:
for Gd, 1 octet, 48 sextet, 120 quartet and 113 doublet states,
for Tb, 7 septet, 140 quintet, 113 triplet and 123 singlet states,
for Dy, 21 sextet, 128 quartet and 130 doublet states,
for Ho, 35 quintet, 210 triplet and 196 singlet states, and 
for Er, 35 quartet and 112 doublets states.
As the basis set for the calculations, ANO-RCC was used. 
The contraction of the basis set is shown in Table \ref{ano-rcc}.
The Cholesky decomposition threshold was set to $5 \times 10^{-8}$ Hartree. 
The obtained SO-RASSI wave functions were transformed into pseudo spin states (or pseudo $\tilde{J}$ states)
\cite{Chibotaru2012SM, Chibotaru2013SM, Ungur2015SM, Chibotaru2015SM}
to analyze the magnetic data using SINGLE\_ANISO module \cite{single_anisoSM}.

\begin{table}[tb]
\begin{ruledtabular}
\caption{Contractions of the employed ANO-RCC basis sets for the {\it ab initio} calculations.}
\label{ano-rcc}
\begin{tabular}{llll}
Ln & 7s6p4d2f1g & Si & 4s3p\\
La & 7s6p4d2f   & O  & 3s2p\\
N (N$_2$ bridge) & 3s2p1d  & C  & 3s2p\\
N (the others)   & 3s2p    & H  & 2s\\
\end{tabular}
\end{ruledtabular}
\end{table}

The obtained crystal-field (CF) levels are shown in Table \ref{cf}. 
In all cases, the lowest spin-orbit states are doubly degenerate (Kramers doublet for Ln $=$ Gd, Dy, Er) 
or quasidegenerate (Ising doublet for Ln $=$ Tb, Ho).
The ground CF states $|\psi\rangle$ are decomposed into the sum of the 
ground pseudo $\tilde{J}$ multiplets $|JM\rangle$ \cite{Ungur2015SM, Chibotaru2015SM}:
\begin{eqnarray}
 |\psi\rangle = \sum_{M=-J}^J C_M |JM\rangle.
\end{eqnarray}
The coefficients $C_M$ are shown in Table \ref{cf_state}.
The contributions of the multiplets with the largest projection ($|M|=J$) to the ground CF states
are 94.2 \%, 96.4 \%, 97.1 \%,  91.7 \%, 78.6 \%, for Gd, Tb, Dy, Ho, and Er, respectively.
For each ground doublets, the $g$-tensors are calculated (Table \ref{gtensor}).
The Er complex is not magnetically anisotropic as much as the other complexes (Tb, Dy, Ho). 
This is because the multiplets $|JM\rangle$ with small $M$ ($|M| < J$) are mixed more than the other systems.

\begin{table}[tb]
\begin{ruledtabular}
\caption{The lowest spin-orbit levels of Ln centers obtained by {\it ab initio} fragment calculations (cm$^{-1}$).
The covalency effect is not included.}
\label{cf}
\begin{tabular}{ccccc}
 Gd & Tb & Dy & Ho & Er \\
\hline
0.000 &   0.000 &   0.000 &   0.000 &   0.000 \\
0.000 &   0.099 &   0.000 &   0.982 &   0.000 \\
0.329 & 141.153 & 179.143 &  87.999 &  74.691 \\
0.329 & 142.222 & 179.143 &  88.534 &  74.691 \\
0.631 & 288.590 & 320.747 & 130.818 & 118.034 \\
0.631 & 295.694 & 320.747 & 147.454 & 118.034 \\
1.108 & 401.446 & 406.717 & 167.052 & 166.279 \\
1.108 & 435.925 & 406.717 & 202.496 & 166.279 \\
      & 490.539 & 470.573 & 224.559 & 212.344 \\
      & 531.201 & 470.573 & 241.625 & 212.344 \\
      & 547.372 & 531.942 & 246.712 & 262.700 \\
      & 730.715 & 531.942 & 284.704 & 262.700 \\
      & 731.087 & 623.187 & 296.344 & 295.345 \\
      &         & 623.187 & 323.988 & 295.345 \\
      &         & 749.919 & 327.012 & 396.290 \\
      &         & 749.919 & 385.730 & 396.290 \\
      &         &         & 386.659 &         \\
\end{tabular}
\end{ruledtabular}
\end{table}

\begin{table}[tb]
\begin{ruledtabular}
\caption{$|JM\rangle$ structure of ground CF doublet on Ln$^{3+}$ center in {\bf 1}-{\bf 5}}
\label{cf_state}
\begin{tabular}{cccccccccc}
 \multicolumn{2}{c}{Gd} & \multicolumn{2}{c}{Tb} & \multicolumn{2}{c}{Dy} & \multicolumn{2}{c}{Ho} & \multicolumn{2}{c}{Er} \\
 $M$ & $|C_M|$          & $M$ & $|C_M|$          & $M$ & $|C_M|$          & $M$ & $|C_M|$          & $M$ & $|C_M|$          \\
\hline
 $-7/2$ & 0.971         & $-6$ & 0.694           & $-15/2$ & 0.986        & $-8$ & 0.677           & $-15/2$ & 0.887 \\
 $-5/2$ & 0.001         & $-5$ & 0.005           & $-13/2$ & 0.019        & $-7$ & 0.005           & $-13/2$ & 0.112 \\
 $-3/2$ & 0.225         & $-4$ & 0.123           & $-11/2$ & 0.164        & $-6$ & 0.162           & $-11/2$ & 0.321 \\
 $-1/2$ & 0.004         & $-3$ & 0.014           & $ -9/2$ & 0.027        & $-5$ & 0.044           & $ -9/2$ & 0.168 \\
 $ 1/2$ & 0.077         & $-2$ & 0.024           & $ -7/2$ & 0.023        & $-4$ & 0.084           & $ -7/2$ & 0.214 \\
 $ 3/2$ & 0.002         & $-1$ & 0.008           & $ -5/2$ & 0.007        & $-3$ & 0.056           & $ -5/2$ & 0.103 \\
 $ 5/2$ & 0.038         &   0  & 0.009           & $ -3/2$ & 0.010        & $-2$ & 0.039           & $ -3/2$ & 0.105 \\
 $ 7/2$ & 0.000         &   1  & 0.008           & $ -1/2$ & 0.004        & $-1$ & 0.033           & $ -1/2$ & 0.024 \\
        &               &   2  & 0.024           &    1/2  & 0.002        &   0  & 0.027           &    1/2  & 0.031 \\
        &               &   3  & 0.014           &    3/2  & 0.001        &   1  & 0.033           &    3/2  & 0.021 \\
        &               &   4  & 0.123           &    5/2  & 0.001        &   2  & 0.039           &    5/2  & 0.011 \\
        &               &   5  & 0.005           &    7/2  & 0.000        &   3  & 0.056           &    7/2  & 0.012 \\
        &               &   6  & 0.694           &    9/2  & 0.000        &   4  & 0.084           &    9/2  & 0.016 \\
        &               &      &                 &   11/2  & 0.000        &   5  & 0.044           &   11/2  & 0.002 \\
        &               &      &                 &   13/2  & 0.000        &   6  & 0.162           &   13/2  & 0.005 \\
        &               &      &                 &   15/2  & 0.000        &   7  & 0.005           &   15/2  & 0.000 \\
        &               &      &                 &         &              &   8  & 0.677           & \\
\end{tabular}
\end{ruledtabular}
\end{table}

\begin{table}[bt]
\begin{ruledtabular}
\caption{The $g$ tensors for the lowest doublets of Ln centers obtained from the fragment calculations.
The transverse $g$-factors for Tb and Ho are zero because of the Griffith's theorem \cite{Griffith1963SM}.
}
\label{gtensor}
\begin{tabular}{cccccc}
      & Gd & Tb & Dy & Ho & Er \\
\hline
$g_X$ &  0.492 &  0.000 &  0.0026 &  0.000 &  0.163 \\
$g_Y$ &  0.824 &  0.000 &  0.0040 &  0.000 &  0.227 \\
$g_Z$ & 13.439 & 17.675 & 19.6459 & 19.422 & 16.528 \\
\end{tabular}
\end{ruledtabular}
\end{table}

\subsection{Calculation of atomic $J$-multiplets of Ln$^{2+}$ ions}
\label{Sec:Ln2+}
The excitation energies of the intermediate virtual electron transferred states were 
replaced by the excitation energies for isolated Ln$^{2+}$ ion (Ln $=$ Gd, Tb, Dy, Ho, Er).
To obtain the energies, the CASSCF and the SO-RASSI calculations were performed with ANO-RCC QZP basis set 
\cite{molcasSM}.
As in the case of the fragment calculations, 
all $4f$ orbitals are treated as the active orbitals of the CASSCF calculations.
In the SO-RASSI calculations, the following $LS$ terms are included: 
${}^7F$ for Gd$^{2+}$, 
${}^6P$, ${}^6F$, ${}^6H$ for Tb$^{2+}$, 
${}^5D$, ${}^5F$, ${}^5G$, ${}^5I$ for Dy$^{2+}$, 
${}^4F$, ${}^4G$, ${}^4I$ for Ho$^{2+}$, and ${}^3F$, ${}^3H$ for Er$^{2+}$.
The excitation energies $\Delta E$ are shown in Table \ref{Ln2+SM}.

\begin{table*}[tb]
\begin{ruledtabular}
\caption{Excitation energies with respect to the lowest $J$ multiplet 
of isolated Ln$^{2+}$ ions (meV).}
\label{Ln2+SM}
\begin{tabular}{ccccccccc}
 \multicolumn{3}{c}{Gd} & \multicolumn{3}{c}{Tb} & \multicolumn{3}{c}{Dy}
\\
$LS$ term & $J$ & $\Delta E$ & $LS$ term & $J$ & $\Delta E$ & $LS$ term & $J$ & $\Delta E$ \\
\hline
${}^7F$ & 6 &   0.000 & ${}^6H$ & 15/2 &    0.000 & ${}^5I$ & 8 &    0.000 \\
        & 5 & 182.104 &         & 13/2 &  307.374 &         & 7 &  458.212 \\
        & 4 & 333.856 &         & 11/2 &  573.765 &         & 6 &  859.148 \\
        & 3 & 455.259 &         &  9/2 &  799.172 &         & 5 & 1202.808 \\
        & 2 & 546.311 &         &  7/2 &  983.596 &         & 4 & 1489.190 \\
        & 1 & 607.012 &         &  5/2 & 1127.038 & ${}^5G$ & 6 & 3412.346 \\
        & 0 & 637.362 & ${}^6F$ & 11/2 & 1050.937 &         & 5 & 3756.005 \\
        &   &         &         &  9/2 & 1276.345 &         & 4 & 4042.388 \\
        &   &         &         &  7/2 & 1460.769 & ${}^5F$ & 5 & 2369.357 \\
        &   &         &         &  5/2 & 1604.210 &         & 4 & 2655.740 \\
        &   &         & ${}^6P$ &  7/2 & 4357.932 & ${}^5D$ & 4 & 5667.150 \\
        &   &         &         &  5/2 & 4501.373 & \\
\hline
 \multicolumn{3}{c}{Ho} & \multicolumn{3}{c}{Er} \\
$LS$ term & $J$ & $\Delta E$ & $LS$ term & $J$ & $\Delta E$ \\
\hline
${}^4I$ & 15/2 &    0.000 & ${}^3H$ & 6 &    0.000 \\
        & 13/2 &  638.260 &         & 5 &  850.945 \\
        & 11/2 & 1191.419 &         & 4 & 1197.911 \\
        &  9/2 & 1659.477 & ${}^3F$ & 4 & 1560.065 \\
${}^4G$ & 11/2 & 3521.812 & \\
        &  9/2 & 3989.870 & \\
${}^4F$ &  9/2 & 2461.643 & \\
\end{tabular}
\end{ruledtabular}
\end{table*}

\begin{table}[tb]
\begin{ruledtabular}
\caption{Kinetic contributions to the CF parameters $\mathcal{J}_{kq00}$ (cm$^{-1}$) for complexes {\bf 1}-{\bf 5}.}
\label{Table_JSM}
\begin{tabular}{ccccccc}
$k$ & $q$     & \multicolumn{5}{c}{$\mathcal{J}_{kq00}$}\\
    &         & Gd & Tb & Dy & Ho & Er \\
\hline
0 & 0            & $-94.88$               & $-95.77$                 & $-70.78$               & $-55.38$ & $-24.20$ \\
4 & 0            &  5.86 $\times 10^{-3}$ & $-30.11$                 &   23.76                &   10.00  & $-11.15$ \\  
4 & $\pm$4       &  3.50 $\times 10^{-3}$ & $-18.00$                 &   14.20                &    5.97  &  $-6.66$ \\
6 & 0            &  4.12 $\times 10^{-7}$ & $-4.95 \times 10^{-1}$   &    6.13                &  $-8.53$ &    4.29  \\
6 & $\pm$4       & $-7.70 \times 10^{-7}$ & $ 9.27 \times 10^{-1}$   & $-11.46$               &   15.96  &  $-8.02$ \\
\end{tabular}
\end{ruledtabular}
\end{table}

\begin{table}[tb]
\begin{ruledtabular}
\caption{Energies of the CF multiplets (cm$^{-1}$) of Ln centers in {\bf 1}-{\bf 5} originating from the ground atomic $J$-multiplet of the corresponding Ln$^{3+}$ ions, calculated with included kinetic contribution.
Due to the latter, the ground CF multiplets of Ln centers are stabilized by 
95, 110, 58, 55, 14 cm$^{-1}$ for Gd, Tb, Dy, Ho, Er, respectively.}
\label{Table_covalent}
\begin{tabular}{ccccc}
Gd & Tb & Dy & Ho & Er \\
\hline
0.000 &   0.000 &   0.000 &   0.000 &   0.000 \\
0.000 &   0.055 &   0.000 &   1.297 &   0.000 \\
0.329 & 168.191 & 163.450 &  95.958 &  68.276 \\
0.329 & 168.965 & 163.450 &  96.064 &  68.276 \\
0.630 & 316.136 & 300.460 & 129.220 & 112.297 \\
0.630 & 318.418 & 300.460 & 147.698 & 112.297 \\
1.108 & 426.550 & 396.451 & 174.642 & 150.837 \\
1.108 & 444.805 & 396.451 & 211.117 & 150.837 \\
      & 501.172 & 458.147 & 223.439 & 197.730 \\
      & 541.183 & 458.147 & 248.194 & 197.730 \\
      & 556.192 & 510.902 & 250.839 & 249.831 \\
      & 740.636 & 510.902 & 280.160 & 249.831 \\
      & 741.002 & 604.121 & 290.136 & 276.581 \\
      &         & 604.121 & 320.445 & 276.581 \\
      &         & 747.503 & 322.767 & 387.217 \\
      &         & 747.503 & 371.677 & 387.217 \\
      &         &         & 371.948 & \\
\end{tabular}
\end{ruledtabular}
\end{table}

\begin{table}[tb]
\begin{ruledtabular}
\caption{Energy of the low-lying exchange KDs (cm$^{-1}$) in {\bf 1}-{\bf 5}}
\label{Table_exchange}
\begin{tabular}{ccccc}
 Gd & Tb & Dy & Ho & Er \\
\hline
 0.000 &   0.000 &   0.000 &   0.000 &   0.000 \\
 0.381 & 207.619 & 120.686 & 105.154 &  27.999  \\
 0.643 & 207.670 & 120.686 & 106.791 &  28.000  \\
 0.865 & 210.623 & 158.882 & 108.718 &  53.467  \\
 1.187 & 227.323 & 164.275 & 110.590 &  64.209  \\
 1.605 & 362.446 & 252.462 & 146.181 &  68.710  \\
 2.112 & 366.170 & 273.941 & 153.019 &  86.241  \\
27.527 & 369.751 & 273.943 & 160.514 &  86.254 \\
27.761 & 369.876 & 293.589 & 161.003 &  99.987 \\
\end{tabular}
\end{ruledtabular}
\end{table}

\section{Analysis of first rank exchange parameters}
\label{Sec:exchangeH}
As shown in Table II in the main text, the first rank part $(k=k'=1)$ of the exchange interaction 
is isotropic Heisenberg type in all complexes, i.e.,
\begin{eqnarray}
 \mathcal{J}_{1\pm 1 1\mp 1} = -\mathcal{J}_{1010} \ne 0,
\label{Eq:isoHeis}
\end{eqnarray}
and the other $\mathcal{J}_{1q1q'}$ are zero.
The reason can be understood analyzing the formula of the exchange interaction.
The exchange parameter between $J$ multiplet and isotropic spin $1/2$ (Eqs. (2), (3) in the main text) is written as 
\cite{Iwahara2015SM}
\begin{eqnarray}
 \mathcal{J}_{kqk'q'} &=& \sum_x \sum_{\alpha_J J} 
 \frac{\left\{t \times t \right\}_{kqk'q'}^x \mathcal{G}^1_{\alpha_J J k' x k} \tilde{\mathcal{F}}^2_{k'}}
  {U_0 + \Delta E_{\alpha_J J}^{n+1}},
\label{Eq:J}
\end{eqnarray}
where 
\begin{eqnarray}
 \{t \times t\}_{kqk'q'}^x &=& (-1)^{l_1-k'+q'} 
\sum_{mm'} \sum_\xi 
\nonumber\\
 &\times&
t_{m\pi^*}^{12} t_{\pi^*m'}^{21}
 C_{l_1m'kq}^{x\xi}C_{k'-q'l_1m}^{x\xi},
\label{Eq:tt}
\end{eqnarray}
$t_{m\pi^*}$ is the electron transfer between the $4f$ with component $m$ of orbital angular momentum
and the $\pi^*$ orbital of N$_2$, 
$l_1 = 3$ is the magnitude of the atomic orbital angular momentum for $f$ orbital,
$x$ $(l_1 - k' \le x \le l_1 + k')$ indicates a rank, $\xi = -x, -x + 1, ..., x$,
$C_{l_1m'kq}^{x\xi}$ and $C_{k'-q'l_1m}^{x\xi}$ are Clebsch-Gordan coefficients \cite{Varshalovich1988SM}, 
$\alpha_J$ and $J$ are the $LS$-term and the total angular momentum of Ln$^{2+}$, respectively,
$\Delta E_{\alpha_J J}^{n+1}$ is the excitation multiplet energies of Ln$^{2+}$, and
$\mathcal{G}^\text{Ln}_{\alpha_J J k' x k}$ and $\tilde{\mathcal{F}}^{\text{N}_2}_{k'}$ are functions of their subscripts.
For the detailed description of $x$, $\mathcal{G}^\text{Ln}_{\alpha_J J k' x k}$, and $\tilde{\mathcal{F}}^{\text{N}_2}_{k'}$,
see Ref. \onlinecite{Iwahara2015SM}.

Since the dependence of the exchange parameter (\ref{Eq:J}) on $q$ and $q'$ appears only in 
$\{t\times t\}_{kqk'q'}^x$ (\ref{Eq:tt}), the condition for the isotropy of $\mathcal{J}_{kq1q'}$ is revealed from the equation.
First, we consider the cases where only the transfer between $f_{\pm m_0}$ orbitals ($m_0 = 0, 1, 2, 3$) 
and the isotropic spin is nonzero for simplicity.
The values of $\{t\times t\}_{1q1q'}^x$ are tabulated in Table \ref{Table:tt}.
We find that the condition (\ref{Eq:isoHeis}) is fulfilled when $m_0 = 2$, while it is not for other $m_0$.
In the case of $m_0 = 1$, the nonzero terms with $q = q' = \pm 1$ are also the source of the anisotropic exchange.
When more than one set of $f$ orbitals $m_0$ contribute to the electron transfer, 
the exchange interaction becomes always anisotropic.
Finally, since Eq. (\ref{Eq:tt}) is independent of ions, 
the condition given above applies to the exchange interaction between any $f$ electron ions and spin $1/2$.

\begin{table}[bt]
\caption{$\{t \times t\}_{1q1q'}^x$ for $m_0 = 0, 1, 2, 3$.}
\label{Table:tt}
\begin{ruledtabular}
\begin{tabular}{ccccccc}
$x$& $q_1$ & $q_2$ & \multicolumn{4}{c}{$m_0$} \\
   &       &       &  0 & 1 & 2 & 3 \\
\hline
2 & 0     & 0     & $\frac{3}{7}|t^{12}_{0\pi^*}|^2$ 
                  & $\frac{16}{21}|t^{12}_{1\pi^*}|^2$
                  & $\frac{10}{21}|t^{12}_{2\pi^*}|^2$
                  & 0\\
2 &$\pm 1$&$\mp 1$& $-\frac{1}{7}|t^{12}_{0\pi^*}|^2$ 
                  & $-\frac{1}{3}|t^{12}_{1\pi^*}|^2$ 
                  & $-\frac{10}{21}|t^{12}_{2\pi^*}|^2$
                  & $-\frac{5}{7}|t^{12}_{3\pi^*}|^2$\\
2 &$\pm 1$&$\pm 1$& 0                                 
                  & $-\frac{2}{7}(t^{12}_{\pm 1\pi^*})^2$ 
                  & 0
                  & 0\\
3 & 0     & 0     & 0                                 
                  & $-\frac{1}{6}|t^{12}_{1\pi^*}|^2$
                  & $-\frac{2}{3}|t^{12}_{2\pi^*}|^2$
                  & $-\frac{3}{2}|t^{12}_{3\pi^*}|^2$\\
3 &$\pm 1$&$\mp 1$& $\frac{1}{2}|t^{12}_{0\pi^*}|^2$  
                  & $\frac{11}{12}|t^{12}_{1\pi^*}|^2$
                  & $\frac{2}{3}|t^{12}_{2\pi^*}|^2$
                  & $\frac{1}{4}|t^{12}_{3\pi^*}|^2$\\
3 &$\pm 1$&$\pm 1$& 0                                 
                  & $-\frac{1}{2}(t^{12}_{\pm 1\pi^*})^2$
                  & 0
                  & 0\\
4 & 0     & 0     & $\frac{4}{7}|t^{12}_{0\pi^*}|^2$  
                  & $\frac{15}{14}|t^{12}_{1\pi^*}|^2$
                  & $\frac{6}{7}|t^{12}_{2\pi^*}|^2$
                  & $\frac{1}{2}|t^{12}_{3\pi^*}|^2$\\
4 &$\pm 1$&$\mp 1$& $-\frac{5}{14}|t^{12}_{0\pi^*}|^2$ 
                  & $-\frac{3}{4}|t^{12}_{1\pi^*}|^2$
                  & $-\frac{6}{7}|t^{12}_{2\pi^*}|^2$
                  & $-\frac{29}{28}|t^{12}_{3\pi^*}|^2$\\
4 &$\pm 1$&$\pm 1$& 0                                 
                  & $-\frac{3}{14}(t^{12}_{\pm 1\pi^*})^2$ 
                  & 0
                  & 0\\
\end{tabular}
\end{ruledtabular}
\end{table}


%

\end{document}